\documentclass[pdflatex,sn-mathphys-num,iicol]{sn-jnl}

\usepackage{graphicx}%
\usepackage{multirow}%
\usepackage[most]{tcolorbox}
\usepackage{amsmath,amssymb,amsfonts}%
\usepackage{amsthm}%
\usepackage{mathrsfs}%
\usepackage[title]{appendix}%
\usepackage{xcolor}%
\usepackage{textcomp}%
\usepackage{manyfoot}%
\usepackage{booktabs}%
\usepackage{algorithm}%
\usepackage{algorithmicx}%
\usepackage{algpseudocode}%
\usepackage{listings}%
\usepackage{mathtools}

\theoremstyle{thmstyleone}%
\theoremstyle{thmstyletwo}%
\newtheorem{remark}{Remark}%

\theoremstyle{thmstylethree}%
\newtheorem{lemma}{Lemma}%

\definecolor{darkblue}{rgb}{0.0,0.2,0.8}

\newcommand{\trace}{\text{tr}}

\newcommand{\vD}{\mathbf{D}}
\newcommand{\ve}{\mathbf{e}}
\newcommand{\vn}{\mathbf{n}}
\newcommand{\vs}{\mathbf{s}}
\newcommand{\vu}{\mathbf{u}}
\newcommand{\vv}{\mathbf{v}}
\newcommand{\vw}{\mathbf{w}}
\newcommand{\vx}{\mathbf{x}}
\newcommand{\vP}{\mathbf{P}}
\newcommand{\vS}{\mathbf{S}}
\newcommand{\vT}{\mathbf{T}}

\begin{document}

\title[]{A phase-field model of viscous surfaces in fluids}


\author[1]{\fnm{Maximilian} \sur{Kloppe}}\email{Maximilian.Kloppe@math.tu-freiberg.de} 

\author*[1,2,3]{\fnm{Sebastian} \sur{Aland}}\email{Sebastian.Aland@math.tu-freiberg.de}

\affil[1]{\orgdiv{Institute of Numerical Mathematics and Optimization}, \orgname{TU Bergakademie Freiberg}, \orgaddress{\street{Akademiestr. 6}, \city{Freiberg}, \postcode{09599}, \country{Germany}}}

\affil[2]{\orgdiv{Faculty of Computer Science/Mathematics}, \orgname{HTW Dresden}, \orgaddress{\street{Friedrich-List-Platz 1}, \city{Dresden}, \postcode{01069}, \country{Germany}}}

\affil[3]{\orgname{Center for Systems Biology Dresden}, \orgaddress{\street{Pfotenhauerstr. 108}, \city{Dresden}, \postcode{01307}, \country{Germany}}}


\abstract{Viscous surfaces play a crucial role in biological and industrial multiphase flows, yet their simulation remains challenging due to complex rheology and topological changes. We present a thermodynamically consistent phase-field model for viscous surfaces in incompressible Navier-Stokes fluids, incorporating surface shear and dilatational viscosities via the Boussinesq-Scriven constitutive law. The diffuse-interface formulation is derived from a sharp-interface model and rigorously validated through matched asymptotic analysis, confirming recovery of the correct sharp-interface limit. We develop an unconditionally energy-dissipative time discretization, ensuring numerical stability. The model is validated through a quantitative benchmark against analytical solutions for surface flows on a rigid sphere, and a simulation of vesicle fusion showing preservation of energy dissipation across topological transitions. These results demonstrate that the framework enables robust, physically consistent simulations of complex surface dynamics, including large deformations and topological changes, without explicit surface tracking.
}




\maketitle

\section{Introduction}\label{sec:Intro}
Viscous surfaces are ubiquitous and play an important role in many biological and multiphase flow systems, ranging from lipid bilayer membranes~\cite{den2007intermonolayer} and biological capsules~\cite{LiZhang2021} to surfactant-laden fluid interfaces~\cite{LOPEZ1998231} and particle-laden interfaces~\cite{MENDOZA2014303}. 

Surface viscosity accounts for the resistance of an interface to in-plane deformation and therefore introduces additional interfacial stresses beyond those associated with surface tension~\cite{Sagis2011}. 
This viscous response is characterized by two material parameters: the surface shear viscosity $\eta_s$, which quantifies the resistance to tangential shearing, and the surface dilatational (bulk) viscosity $\eta_b$, which characterizes the resistance to in-plane expansion and compression. These contributions to the interfacial stress, first formulated by Scriven~\cite{scriven1960dynamics}, can significantly influence the dynamics of the surrounding fluid. 
For a characteristic system of size $R$, this regime is characterized by $\eta_s$ or $\eta_b$ being comparable or larger than the bulk fluid viscosity at that scale, i.e., $\eta_s,\eta_b \not\ll R(\eta_0+\eta_1)$.

Current numerical simulation approaches commonly rely on sharp-interface methods to account for viscous or viscoelastic surface rheology~\cite{Barrett2015,WITTWER2023100126,DEKINKELDER2021110413}. Typically, the surface is represented by an explicitly tracked mesh that evolves together with the surrounding fluid. While this provides an accurate representation of the surface geometry and rheology, such methods are often limited  when dealing with strong deformations or topological changes. Large deformations can lead to severe distortion of the surface mesh and require remeshing or other mesh-management techniques to maintain a suitable discretization~\cite{sahu2020arbitrary}. Topological changes, such as the merging or breakup of surfaces, cannot be captured by continuous deformation of the existing mesh and therefore require explicit modifications of the mesh connectivity or a reconstruction of the surface~\cite{kovacs2024numerical}.

In this work we propose an alternative formulation based on a diffuse interface description of the surface by a phase-field, which naturally allows topological changes. Depending on the application of interest, phase-field methods additionally offer some distinct advantages over other interface capturing methods. 
In particular, they can intrinsically include mass conservation and transport stabilization, while also providing a framework for fully discrete energy-stable schemes, see e.g. \cite{shen2018scalar, aland2016efficient}, for two-phase flows. 
Moreover, a linearly implicit coupling between phase-field advection and the flow equations can avoid restrictive time-step constraints arising from stiff interfacial forces~\cite{ALAND2014_TimeIntegration}. 
The diffuse-interface description also naturally regularizes the contact-line singularity, making phase-field methods particularly suitable for wetting problems~\cite{qian2006variational, demont2026consistency, AlandMokbel2021, Mokbel2024Wetting}.

Recently, phase-field formulations have been extended to account for fluid-structure interaction~\cite{MOKBEL2018823} and interfacial in-plane elasticity~\cite{KLOPPE2024}. 
The in-plane viscous response to surface compression and shearing, however, has received comparatively little attention in this framework. One related approach incorporates surface inextensibility~\cite{Aland_2014_inextensibility}, which can be interpreted as the limiting case of infinite dilatational viscosity.

More recently, Park~\cite{Park2026} introduced a diffuse-interface formulation incorporating Boussinesq--Scriven surface stresses into a Stokes--Cahn--Hilliard model, yet without a rigorous derivation, verification, or validation.

In the following, we introduce a rigorous formulation and validation of surface viscosity in phase-field fluid models. 
We first phrase a sharp interface description of viscous surface stresses into a thermodynamically consistent diffuse interface formulation (Secs.~\ref{sec:sharp interface model}--\ref{sec:energy stability model}). 
We then recover the corresponding correct sharp interface limit by means of an asymptotic analysis (Sec.~\ref{sec:matched asymp analysis}). 
Based on the resulting formulation, we derive an energy-stable numerical scheme (Sec.~\ref{sec:discretization}) and validate the model in a series of numerical tests (Secs.~\ref{sec: viscous band} and \ref{sec:validation rigid sphere}). Finally, we demonstrate the capability of the framework to handle viscous surfaces undergoing topological changes (Sec.~\ref{sec:topological change}). As a result, we obtain a rigorous framework for surface viscosity in phase-field fluid models that extends the advantages of diffuse-interface methods to complex surface rheology.

\section{Model}\label{sec:Model}
\subsection{Sharp Interface Model of a Viscous Surface}\label{sec:sharp interface model}
Throughout this work we see the following sharp interface setting as the ground truth to describe a two-phase flow problem of a viscous surface and aim to approximate it using a diffuse interface approach. 

\begin{figure*}
    \centering
    \includegraphics[width=0.99\linewidth]{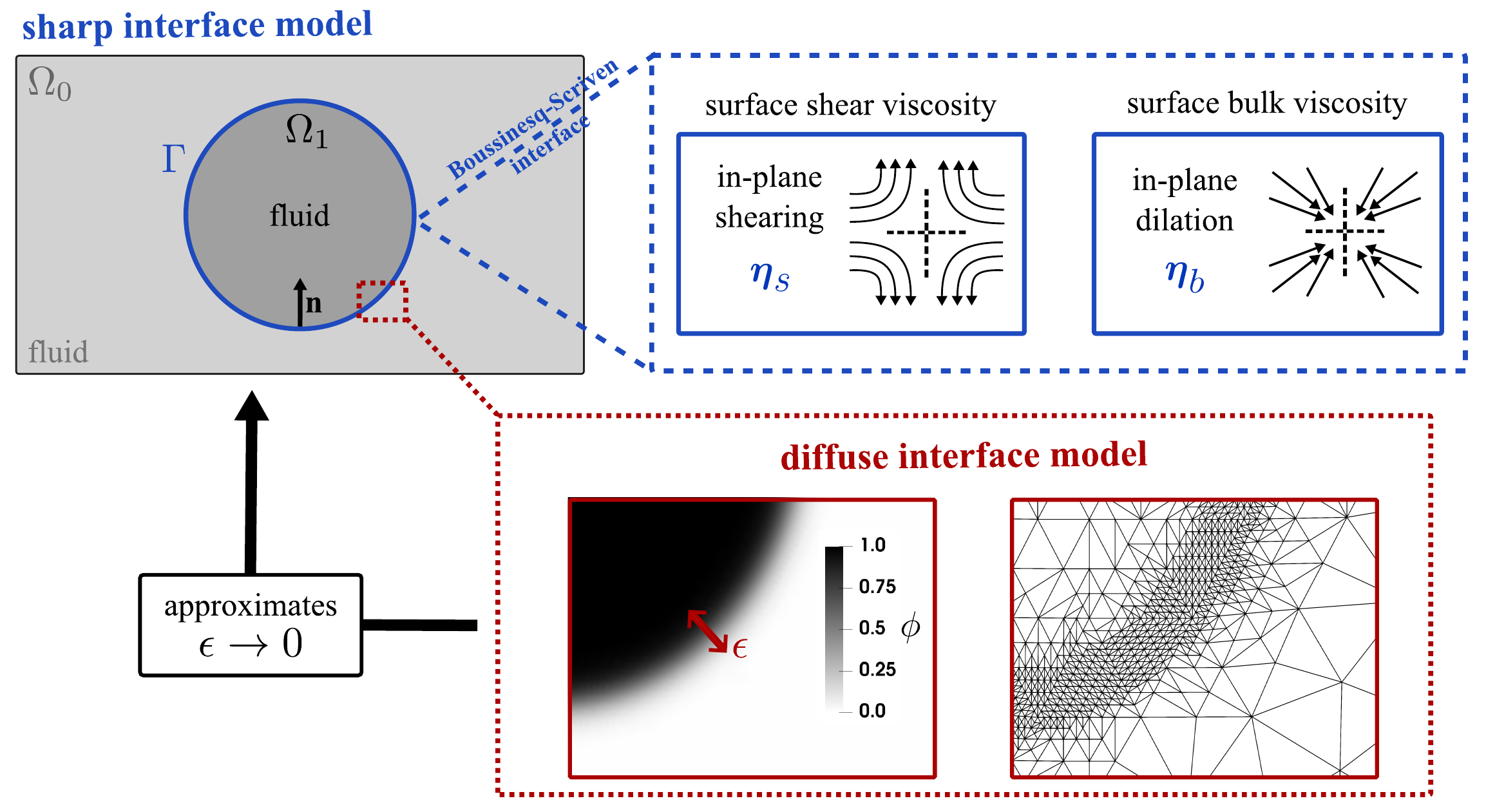}
    \caption{\textbf{Top-Left:} Illustration of the domain $\Omega = \Omega_0 \cup \Omega_1 \cup \Gamma$ for a two-phase flow problem separated by a sharp viscous fluid interface $\Gamma$. \textbf{Top-Right:} Schematic representation of the Boussinesq-Scriven surface fluid model, characterized by two distinct viscous moduli. $\boldsymbol\eta_s$: Surface shear  viscosity $\eta_s$ penalizes the rate of area-preserving deformations of the interface, e.g. a purely extensional surface flow. $\boldsymbol\eta_b$: Surface dilational viscosity $\eta_b$ penalizes the rate of area change, e.g. surface compressive flows. \textbf{Bottom:} Diffuse interface representation of the inset region by a phase-field $\phi$ with interface width $\epsilon$ (left) and the
    corresponding adaptive mesh (right)}
    \label{fig:sharp interface setup}
\end{figure*}

We consider a time-dependent closed interface $\Gamma$ within a domain $\Omega \subset \mathbb{R}^d$ ($d=2,3$), separating it into two subdomains $\Omega_0$ and $\Omega_1$, representing the outer and inner fluids, respectively (see Fig.~\ref{fig:sharp interface setup}). The flow of the immiscible fluids is governed by the incompressible Navier--Stokes equations in the bulk phases $\Omega_i, \, i=0,1$:
\begin{align} \label{eq: ns 1}
    \rho_i (\partial_t \mathbf{v} + \mathbf{v} \cdot \nabla \mathbf{v})  - \nabla \cdot \left(2\eta_i \mathbf{D}(\mathbf{v})\right) + \nabla p_i &= 0 \\ 
    \nabla \cdot \mathbf{v} &= 0 
\end{align}
where $\mathbf{v}$ is a continuous velocity field to indicate the movement of the fluids and by $ \mathbf{D}(\mathbf{v})= \frac{1}{2} (\nabla \mathbf{v} + (\nabla\mathbf{v})^T)$ we denote the rate of deformation tensor. Also $\rho_i, \eta_i, p_i$ ($i=0,1$) denote the density, viscosity and pressure in the two fluids, respectively.

At the interface $\Gamma$, the force balance requires that the jump in the bulk fluid stress is balanced by intrinsic surface forces. We consider a viscous interface governed by the Boussinesq-Scriven constitutive law~\cite{scriven1960dynamics}. The interface condition reads:
\begin{equation} \label{eq: sharp interface condition}
    \left[ -p_i \mathbf{I} + 2 \eta_i \mathbf{D}(\mathbf{v}) \right]_0^1 \cdot \mathbf{n} = \sigma \kappa \mathbf{n} + \nabla_\Gamma \cdot \mathbf{S}_\Gamma \quad \text{on } \Gamma,
\end{equation}
with constant surface tension $\sigma$ and total curvature $\kappa = \nabla_\Gamma \cdot \mathbf{n}$. Here, $\mathbf{n}$ is the interface normal vector pointing into $\Omega_1$, $[\cdot]_0^1 = [\cdot]_1 -[\cdot]_0$ denotes the jump of a quantity across the interface,  $\nabla_\Gamma$ is the surface gradient, and  $\nabla_\Gamma \cdot$ the surface divergence respectively. The surface stress tensor $\mathbf{S}_{\Gamma}$ is defined using the surface projection tensor $\mathbf{P}_\Gamma = \mathbf{I} - \mathbf{n} \otimes \mathbf{n}$ as:
\begin{equation}\label{eq:sharp_stress_tensor}
    \mathbf{S}_{\Gamma} = (\eta_b - \eta_s) (\nabla_\Gamma \cdot \mathbf{v}) \mathbf{P}_\Gamma + 2 \eta_s \mathbf{D}_\Gamma ,
\end{equation} 
with the surface rate of deformation tensor $\mathbf{D}_\Gamma = \frac{1}{2} \mathbf{P}_\Gamma (\nabla \mathbf{v} + (\nabla \mathbf{v})^T) \mathbf{P}_\Gamma$. Here, $\eta_s \ge 0$ represents the surface shear viscosity and $\eta_b \ge 0$ represents the surface bulk (dilational) viscosity. The effect of both parameters is also displayed in Fig.~\ref{fig:sharp interface setup}.

The resulting total energy of this system consists of the kinetic energy $E_{kin}$ and the surface free energy $E_\sigma$ and reads
\begin{equation}
    E_{tot}(t) = E_{kin}(t) + E_{\sigma}(t) = \int_\Omega \frac{1}{2}\rho \vert\mathbf{v}\vert^2 \, d\mathbf{x} + \int_\Gamma \sigma \, d\mathbf{s}.
\end{equation} 
It can be shown that energy is dissipated under evolution of the system \eqref{eq: ns 1}-\eqref{eq:sharp_stress_tensor},

\begin{align}
    d_t E_{tot} =& - \int_{\Omega \setminus \Gamma(t)} 2\eta \Vert \mathbf{D}(\mathbf{v}) \Vert_F^2 \, d\mathbf{x} \nonumber\\
    &- \int_{\Gamma(t)} \left(\eta_b + (3-d) \eta_s \right)(\nabla_\Gamma \cdot \mathbf{v})^2 d\mathbf{s} \nonumber\\
     &- \int_{\Gamma(t)} 2\eta_s \Vert\bar{\mathbf{D}}_\Gamma\Vert_F^2 d\mathbf{s} \nonumber \\
    \le & 0, \label{eq:sharp interface energy time derivative decreasing}
\end{align} 
where $d=2, 3$, $\Vert\cdot \Vert_F$ denotes the Frobenius norm, and $\bar{\mathbf{D}}_\Gamma$ the deviatoric part of the surface rate of deformation tensor 
\begin{equation}
    \mathbf D_\Gamma
    =
    \bar{\mathbf D}_\Gamma
    +
    \frac{1}{d-1}
    (\nabla_\Gamma\cdot\mathbf v)\mathbf P_\Gamma,
\end{equation}
see \cite{Barrett2015,bothe2010}.

\begin{remark} \label{rm: remark on S}
In the case $d=2$, $\Gamma$ is a one-dimensional curve and the deviatoric-dilational decomposition becomes trivial as the deviatoric part vanishes, i.e.,
\begin{equation*}
    \bar{\mathbf D}_\Gamma = \mathbf 0,
\qquad
\mathbf D_\Gamma = (\nabla_\Gamma \cdot \mathbf v)\mathbf P_\Gamma.
\end{equation*}
Hence, the surface stress reduces to
\begin{equation}
    \mathbf S_\Gamma = (\eta_b+\eta_s)(\nabla_\Gamma \cdot \mathbf v)\mathbf P_\Gamma
\end{equation}
and the surface viscosity is governed by a single parameter $\eta_{\rm 1D} = \eta_b+\eta_s$.
\end{remark}

\subsection{Diffuse Interface Formulation of a Viscous Surface} \label{sec: diffuse interface model}
In the above sharp interface model, the interface between the two fluids is considered an evolving free boundary $\Gamma$ that can be tracked explicitly at any time. When using the phase-field method, an auxiliary function $ \phi: \Omega \rightarrow \left[ 0,1\right]  $, the phase-field, is introduced and the interface position is implicitly described by a level-set of $\phi$ instead. In our case this function takes the values $ \phi = 0 $ and $ \phi = 1 $ in the different fluid phases with a smooth transition in between, around the interface. Hence, the interface is diffuse with a finite width $ \epsilon $ and can be located by an intermediate level-set of the phase-field  (here $ \phi = 0.5 $), see Fig. \ref{fig:sharp interface setup}.

The surface free energy can be reformulated into the diffuse interface approximation, the Ginzburg-Landau free energy: \begin{align} 
    \int_\Gamma \sigma \, d\mathbf{s} &\approx E_{GL}(t) \nonumber\\
    &= \int_\Omega \sigma \left( \frac{\epsilon}{2} |\nabla \phi|^2 + \frac{1}{\epsilon} W(\phi) \right) \, d\mathbf{x}, \label{eq: GL energy}
\end{align} where $ W(\phi) $ denotes a double-well potential which we choose as $ W(\phi) = 18 \phi^2 (\phi-1)^2$ to eliminate any scaling factors in this energy.

To incorporate viscous surface rheology into the phase-field formulation, we propose the following extension of the Navier-Stokes Cahn-Hilliard equations in the domain $\Omega$:
\begin{strip}
\vspace{-0.6cm}
\begin{tcolorbox}[title=Phase-Field Viscous Surface Model, colback=gray!10, colframe=gray!50, boxrule=0.5pt ]
\vspace{-0.5cm}
    \begin{align}		
    \rho(\phi)\partial_t \mathbf{v} + \left(\left(\rho(\phi)\mathbf{v} + \mathbf{J}\right) \cdot \nabla\right) \mathbf{v} + \nabla p - \nabla \cdot \left( 2\eta(\phi) \mathbf{D}(\mathbf{v})\right)  &= \sigma \mu \nabla \phi + {\color{darkblue}\nabla \cdot (\vert \nabla \phi \vert \mathbf{S}_{\phi})}, \label{eq: NS balance of momentum} \\
    \nabla \cdot \textbf{v} &= 0, \label{eq: NS balance of mass}  \\
    \partial_t \phi + \mathbf{v} \cdot \nabla \phi &= M \Delta \mu, \label{eq: Cahn Hillard sys 1} \\	
    \mu &= \dfrac{1}{\epsilon} W^\prime (\phi) - \epsilon \Delta\phi, \label{eq: Cahn Hillard sys 2} \\
    \label{eq:diffuse_stress_tensor}
    {\color{darkblue}\mathbf{S}_{\phi}} &{\color{darkblue}= (\eta_b - \eta_s) (\mathbf{P}_\phi \mathbin{:} \nabla \mathbf{v}) \mathbf{P}_\phi + 2 \eta_s \mathbf{D}_\phi}.
\end{align}
\end{tcolorbox}
\end{strip}
Here, the dynamic viscosity $\eta(\phi)$ and the mass density $\rho(\phi)$ are modeled as phase-dependent interpolations between the corresponding bulk fluid properties:
\begin{align*}
    \eta(\phi) &= \eta_0 (1-\phi) + \eta_1 \phi, \\
    \rho(\phi) &= \rho_0 (1- \phi) + \rho_1 \phi.
\end{align*}
Following the thermodynamically consistent framework of Abels, Garcke, and Grün \cite{AGG12}, the velocity $\mathbf{v}$ represents the volume-averaged mixture velocity, which remains solenoidal ($\nabla \cdot \mathbf{v} = 0$) even in the presence of density contrasts. Consequently, the total mass flux in the momentum balance \eqref{eq: NS balance of momentum} is decomposed into a convective contribution $\rho(\phi)\mathbf{v}$ and a relative diffusive mass flux $\mathbf{J} = ({\rho_0 - \rho_1}) M \nabla \mu$ driven by chemical potential gradients. Incorporating $\mathbf{J}$ into the momentum balance accounts for the mass transport relative to the volume-averaged flow, which is strictly required to guarantee thermodynamic consistency and energy dissipation in the variable-density regime. The flux naturally vanishes in the limit of matched densities ($\rho_0 = \rho_1$).

Finally, the tensor $\mathbf{S}_{\phi}$ provides an approximation of the viscous surface stress tensor $\mathbf{S}_{\Gamma}$ to be analyzed and validated in the following sections. 
Here, $\mathbf A:\mathbf B \coloneqq \sum_{i,j=1}^d A_{ij}B_{ij} = \trace(\mathbf{A}^{\text{T}}\mathbf{B})$ denotes the Frobenius inner product for two matrices $\mathbf A,\mathbf B\in\mathbb R^{d\times d}$. 
By introducing the diffuse interface normal $\textbf{n}_\phi = \nabla \phi /\vert \nabla \phi \vert$, we define the corresponding diffuse projection tensor $\mathbf{P}_\phi \coloneqq \mathbf{I} - \textbf{n}_\phi \otimes \textbf{n}_\phi$. Consequently, all surface operators are straightforwardly approximated via this projection: the diffuse surface rate of deformation tensor is given by $\mathbf{D}_\phi \coloneqq \frac{1}{2} \vP_\phi \left( \nabla \textbf{v} +(\nabla \textbf{v})^T \right) \vP_\phi$, the diffuse surface gradient evaluates to $\nabla_{\Gamma,\phi} = \mathbf{P}_\phi \nabla$, and the diffuse surface divergence of the velocity field is naturally recovered as $\nabla_{\Gamma,\phi} \cdot \mathbf{v} = \mathbf{P}_\phi \mathbin{:} \nabla \mathbf{v}$.

\subsection{Thermodynamic consistency}\label{sec:energy stability model}
In accordance with the second law of thermodynamics, an isothermal closed system must evolve towards a state of minimal free energy. In the following, we show that the total energy, consisting of the kinetic energy $E_{kin}$ and the Ginzburg-Landau free energy $E_{GL}$ given in \eqref{eq: GL energy}, decreases over time consistent with the dissipation of the sharp interface model \eqref{eq:sharp interface energy time derivative decreasing}.

\subsubsection{Kinetic energy} \label{subsec: kinetic energy dissipation continuous case}
For the kinetic energy of an incompressible fluid with variable density, it holds
\begin{align*}
    d_t E_{kin} &= d_t \int_\Omega \frac{1}{2}\rho(\phi) \vert\mathbf{v}\vert^2 \, d\mathbf{x} \\
    &= \int_\Omega \left( \frac{1}{2} (\partial_t \rho(\phi)) \vert\mathbf{v}\vert^2 + \rho(\phi) (\partial_t \mathbf{v}) \cdot \mathbf{v} \right) d\mathbf{x}.
\end{align*}
By utilizing the generalized continuity equation $\partial_t \rho(\phi) + \nabla \cdot (\rho(\phi)\vv + \mathbf{J}) = 0$ with the relative mass flux $\mathbf{J} = ({\rho_0 - \rho_1}) M \nabla \mu$, which arises naturally from the linear density interpolation and the Cahn--Hilliard dynamics \cite[cf. Eq. (2.14)]{AGG12}, the first term can be reformulated. Assuming appropriate boundary conditions (e.g., no-slip $\vv\vert_{\partial \Omega} = 0$) such that boundary integrals vanish, applying integration by parts yields
\begin{align*}
    &\int_\Omega \frac{1}{2} (\partial_t \rho(\phi)) \vert\mathbf{v}\vert^2 \, d\mathbf{x} \\
    =&  \int_\Omega -\frac{1}{2} \nabla \cdot (\rho(\phi)\vv + \mathbf{J}) \vert\mathbf{v}\vert^2 \, d\mathbf{x}\\ 
    =& \int_\Omega \left( (\rho(\phi)\vv + \mathbf{J}) \cdot \nabla \vv \right) \cdot \vv \, d\mathbf{x}.
\end{align*}
Recombining both parts, the integrant in the time derivative of the kinetic energy simplifies to exactly match the convective material derivative:
\begin{equation*}
    d_t E_{kin} = \int_\Omega \left( \rho(\phi) \partial_t \vv + \left( (\rho(\phi)\vv + \mathbf{J}) \cdot \nabla \right) \vv \right) \cdot \vv d\mathbf{x}.
\end{equation*}
Inserting Eq.~\eqref{eq: NS balance of momentum}, we obtain
\begin{align*}
    d_t E_{kin} =& \int_\Omega \vv \cdot \left( \nabla \cdot( 2 \eta(\phi) \vD(\vv)) - \nabla p\right) d\mathbf{x} \\
    &+ \int_\Omega \vv \cdot \left(\nabla\cdot \left(\vert \nabla\phi \vert \vS_{\phi}\right) +\sigma \mu \nabla \phi \right) d\mathbf{x}.
\end{align*}
Applying integration by parts and utilizing the incompressibility constraint $\nabla \cdot \vv = 0$, the remaining individual terms evaluate as follows:
\begin{itemize}
    \item pressure: $\int_\Omega \nabla p \cdot \vv \, d\mathbf{x} = - \int_\Omega p (\nabla \cdot \vv) \, d\mathbf{x} = 0$,
    \item dissipation: \begin{align*}
        &\int_\Omega (2 \nabla \cdot( \eta(\phi) \vD(\vv)) + \nabla\cdot \left(\vert \nabla\phi \vert \vS_{\phi}\right) ) \cdot \vv \, d\mathbf{x} \\ =& - \int_\Omega (2 \eta(\phi) \vD(\vv) +\vert \nabla\phi \vert \vS_{\phi}) \mathbin{:} \nabla \vv \, d\mathbf{x}
    \end{align*}.
\end{itemize}
We can conclude that the dissipation is indeed non-positive, since $2 \eta(\phi) \vD(\vv) \mathbin{:} \nabla \vv = 2 \eta(\phi) \Vert \vD(\vv)\Vert_F^2 \ge 0 $ and with Eq.~\eqref{eq:diffuse_stress_tensor} it holds
\begin{align}
    \vS_{\phi} \mathbin{:} \nabla \vv =& (\eta_b - \eta_s) \left((\nabla_{\Gamma, \phi} \cdot \vv) \vP_\phi\right) \mathbin{:} \nabla \vv \nonumber \\
    &+ (2 \eta_s \vP_\phi \vD(\vv) \vP_\phi ) \mathbin{:} \nabla \vv \nonumber \\
    =& (\eta_b - \eta_s) (\nabla_{\Gamma, \phi} \cdot \vv)^2 + 2 \eta_s \Vert \vD_\phi\Vert_F^2. \label{eq: S times nabla v}
\end{align}
After decomposing $\vD_\phi = \bar{\vD}_\phi + \frac{1}{d-1}(\trace(\vD_\phi))\vP_\phi$ into its deviatoric and dilational parts and using that $\Vert \vD_\phi\Vert_F^2 = \Vert \bar{\vD}_\phi \Vert_F^2 +  \frac{1}{d-1}(\trace(\vD_\phi))^2 $ as well as $\trace(\vD_\phi) = \nabla_{\Gamma, \phi} \cdot \vv$ we get
\begin{align} 
    \vS_{\phi} \mathbin{:} \nabla \vv =& \left(\eta_b + (3-d) \eta_s \right)(\nabla_{\Gamma, \phi} \cdot \vv)^2 \nonumber \\ 
    &+ 2 \eta_s \Vert \bar{\vD}_\phi \Vert_F^2 \nonumber \\ 
    \ge& 0 \label{eq: S times nabla v after splitting}
\end{align} for $\eta_b, \eta_s \ge 0$, $d \in \{2, 3\}$.

\subsubsection{Ginzburg Landau energy}
Taking the total time derivative of the Ginzburg-Landau free energy \eqref{eq: GL energy} and applying integration by parts, we obtain:
\begin{equation}
    d_t E_{GL} = \int_\Omega \sigma \mu \partial_t \phi \, d\mathbf{x},
\end{equation}
with chemical potential $\mu$, cf. Eq.~\eqref{eq: Cahn Hillard sys 2}. Also the boundary integral was dropped by assuming ${\nabla \phi \cdot \vn = 0}$. 
Inserting the Cahn-Hilliard equation \eqref{eq: Cahn Hillard sys 1} and applying integration by parts again (assuming $\nabla \mu \cdot \vn = 0$), we get
\begin{equation}
    d_t E_{GL} = - \int_\Omega \sigma M |\nabla \mu|^2 \, d\mathbf{x} - \int_\Omega \sigma \mu \mathbf{v} \cdot \nabla \phi \, d\mathbf{x},
\end{equation}
where the first term represents the chemical dissipation due to diffusion and the second term exactly cancels the surface tension contribution in the kinetic energy balance.

We conclude for the total energy evolution
\begin{align*}
     d_t E_{tot} =& - \int_\Omega \left( 2 \eta(\phi) \Vert \vD(\vv)\Vert_F^2 
     + \sigma M |\nabla \mu|^2 \right) d\mathbf{x} \\
      &- \int_\Omega |\nabla\phi|\left(\eta_b + (3-d) \eta_s \right)(\nabla_{\Gamma, \phi} \cdot \vv)^2  d\mathbf{x} \\ 
     &- \int_\Omega 2 \eta_s |\nabla\phi|\Vert \bar{\vD}_\phi \Vert_F^2  
     d\mathbf{x}     \\ 
     \leq& \, 0, 
\end{align*} 
which proves that the variable-density diffuse-interface system is thermodynamically consistent and dissipative and consistent with the sharp-interface dissipation Eq.~\eqref{eq:sharp interface energy time derivative decreasing}.

\subsection{Matched asymptotic analysis of the viscous surface stress} \label{sec:matched asymp analysis}
Matched asymptotic analysis for the standard two-phase flow Navier--Stokes--Cahn--Hillard model has been done before (see e.g. \cite{AGG12}), establishing the correct sharp interface limit, e.g. for $M \propto \epsilon$, but also other mobility scalings (see e.g. \cite{abels2014sharp,magaletti2013sharp,Demont2023}). Hence, we focus here on the contribution of the viscous surface stress only, where it remains to show that the sharp interface limit recovers Eqs.~\eqref{eq: sharp interface condition} and \eqref{eq:sharp_stress_tensor} and that the velocity $\vv$ remains continuous across $\Gamma$.

Our derivation in the following, is based on the argumentation in \cite{AGG12}.
Let $\Gamma$ denote the $0.5$ level set of $\phi$, and $\vn$ its normal vector pointing into the inner domain $\phi \approx 1$.  First, a new inner coordinate system is introduced in a neighbourhood $U_\epsilon$ of $\Gamma$, such that for any $\vx \in U_\epsilon$ there is a unique $\vs(\vx) \in \Gamma $ with minimal distance to $\vx$. Now one can write 
\begin{equation*}
    \vx = \vs(\vx) + r(\vx) \textbf{n} = s(\vx) + \epsilon z(\vx) \textbf{n},
\end{equation*} with $r(\vx)$ being a signed distance function with $r<0$ in $\Omega_0$ and $r>0$ in $\Omega_1$ and $z(\vx) = \frac{r(\vx)}{\epsilon}$ being a scaled distance function.

Outside of the interface region an outer expansion $\phi = \sum_{i=0}^{\infty} \epsilon^i \phi^{(i)}$ in terms of the original coordinate system is introduced.
Within the neighbourhood of $\Gamma$ we introduce an inner expansion of the phase field function, $\hat{\phi} = \sum_{i=0}^{\infty} \epsilon^i \hat{\phi}^{(i)}$, in terms of the new coordinate system. 
By matching both expansions in an overlapping region, where both are valid representations of the same function, the following matching conditions hold
\begin{align}
    \lim_{r \rightarrow \pm 0 } \phi^{(0)} (\vs,r) &= \lim_{z \rightarrow \pm \infty} \hat{\phi}^{(0)}(\vs,z), \label{eq: matching condition c} \\
    \lim_{r \rightarrow \pm 0 } \nabla \phi^{(0)}(\vs,r) \cdot \textbf{n} &=  \lim_{z  \rightarrow \pm \infty} \partial_z \hat{\phi}^{(1)}(\vs,z). \label{eq: matching cond constant in normal}
\end{align}
Analogously an outer expansion $\vv = \sum_{i=0}^{\infty} \epsilon^i \vv^{(i)}$ and an inner expansion $\hat{\vv} = \sum_{i=0}^{\infty} \epsilon^i \hat{\vv}^{(i)}$ of the velocity field $\vv$ are defined and the matching condition
\begin{equation} \label{eq: matching cond velocities}
    \lim_{r \rightarrow \pm 0 } \vv^{(0)} (\vs,r) = \lim_{z \rightarrow \pm \infty} \hat{\vv}^{(0)}(\vs,z)
\end{equation} holds.

One can easily see, that the outer expansion of the diffuse viscous surface force $\nabla\cdot(|\nabla\phi|\mathbf{S}_\phi)$ becomes zero, since $\vert\nabla \phi\vert = 0$ holds away from the interface. Therefore the term does not interfere with the outer expansion of the classical Navier--Stokes--Cahn--Hilliard equations, thereby recovering the correct sharp interface equations away from $\Gamma$.

\subsubsection{Inner expansion}
For the inner expansion we need to transform the spatial gradient operator into the local coordinates by 
\begin{equation}\label{eq: local gradient}
    \nabla = \nabla_\Gamma + \frac{1}{\epsilon} \vn \partial_z + \mathcal{O}(\epsilon),
\end{equation}
where $\nabla_\Gamma = \mathbf{P}_\Gamma \nabla$ denotes the surface gradient with $\mathbf{P}_\Gamma = \mathbf{I} - \vn \otimes \vn$. 
The analysis in \cite{AGG12} shows that \begin{equation}\label{eq: zero order phi hat}
    \hat{\phi}^{(0)} = \dfrac{1}{2}\left(1 + \tanh{3z} \right)
\end{equation} is independent of $\vs$, hence its tangential gradient vanishes ($\nabla_\Gamma \hat{\phi}^{(0)} = \mathbf{0}$). Thus, we obtain
\begin{equation}
    \nabla \hat{\phi} = \frac{1}{\epsilon} \partial_z \hat{\phi}^{(0)} \vn + \partial_z \hat{\phi}^{(1)} \vn + \mathcal{O}(\epsilon).
\end{equation}
The magnitude of the phase-field gradient expands as
\begin{align}
    |\nabla \hat{\phi}| &= \sqrt{ \frac{1}{\epsilon^2} (\partial_z \hat{\phi}^{(0)})^2 + \frac{2}{\epsilon} \partial_z \hat{\phi}^{(0)} \partial_z \hat{\phi}^{(1)} + \mathcal{O}(1) } \nonumber \\
    &= \frac{1}{\epsilon} |\partial_z \hat{\phi}^{(0)}| +  \partial_z \hat{\phi}^{(1)} + \mathcal{O}(\epsilon), \nonumber 
\end{align} which can be seen from a Taylor expansion of $\sqrt{1+x}$ around $0$ and since $\partial_z \hat{\phi}^{(0)} >0 $.
The diffuse normal vector $\hat{\vn}_\phi = \frac{\nabla \hat{\phi}}{|\nabla \hat{\phi}|}$ is consequently given as
\begin{equation*}
    \hat{\vn}_\phi =  
    \frac{\frac{1}{\epsilon} \partial_z \hat{\phi}^{(0)} + \partial_z \hat{\phi}^{(1)} + \mathcal{O}(\epsilon)}{\frac{1}{\epsilon} |\partial_z \hat{\phi}^{(0)}| + \partial_z \hat{\phi}^{(1)} + \mathcal{O}(\epsilon)} \vn,
\end{equation*}
 yielding $\hat{\vn}_\phi =  \vn + \mathcal{O}(\epsilon^2)$. Consequently, the diffuse projection tensor strictly corresponds to the sharp interface projection tensor in leading order: $\hat{\mathbf{P}}_\phi = \mathbf{I} - \hat{\vn}_\phi \otimes \hat{\vn}_\phi = \mathbf{P}_\Gamma + \mathcal{O}(\epsilon^2)$. 

Next, we evaluate the gradient of the inner velocity field. 
\begin{equation} \label{eq:inner_velocity_grad}
    \nabla \hat{\mathbf{v}} = \frac{1}{\epsilon} \vn \otimes \partial_z \hat{\mathbf{v}}^{(0)} + \nabla_\Gamma \hat{\mathbf{v}}^{(0)} + \vn \otimes \partial_z \hat{\mathbf{v}}^{(1)} + \mathcal{O}(\epsilon).
\end{equation}
We will show below (Sec.~\ref{sec: asymptotic ana v}) that  $\hat{\textbf{v}}^{(0)}$ is constant in $z$-direction, which allows to drop the first term. Substituting Eq.~\eqref{eq:inner_velocity_grad} into the definition of the diffuse stress tensor \eqref{eq:diffuse_stress_tensor}, and utilizing the geometric properties $\mathbf{P}_\Gamma \vn = \mathbf{0}$ and $\vn^T \mathbf{P}_\Gamma = \mathbf{0}$,  the projection operators $\hat{\mathbf{P}}_\phi \approx \mathbf{P}_\Gamma$ filter out the normal derivatives of the first-order perturbation $\hat{\mathbf{v}}^{(1)}$. It holds 
\begin{align*}
    &\hat{\mathbf{P}}_\phi (\nabla \hat{\mathbf{v}}) \hat{\mathbf{P}}_\phi \\
    =& \mathbf{P}_\Gamma \left( \nabla_\Gamma \hat{\mathbf{v}}^{(0)} + \vn \otimes \partial_z \hat{\mathbf{v}}^{(1)} \right) \mathbf{P}_\Gamma + \mathcal{O}(\epsilon) \\
    =& \mathbf{P}_\Gamma \left( \nabla_\Gamma \hat{\mathbf{v}}^{(0)} \right) \mathbf{P}_\Gamma + \mathcal{O}(\epsilon).
\end{align*}
Plugging this result into the definition of the diffuse rate-of-deformation tensor $\mathbf{D}_\phi$, we directly recover the sharp rate-of-deformation tensor $\mathbf{D}_\Gamma$. 

It remains to show that $\hat{\mathbf{P}}_\phi \mathbin{:} \nabla \hat{\mathbf{v}}$ approaches $ \nabla_\Gamma \cdot \hat{\mathbf{v}}^{(0)}$ in the sharp interface limit. 
It holds $\hat{\mathbf{P}}_\phi \mathbin{:} \nabla \hat{\mathbf{v}} = \nabla \cdot (\hat{\mathbf{P}}_\phi \hat{\mathbf{v}}) - (\nabla \cdot \hat{\mathbf{P}}_\phi) \cdot \hat{\mathbf{v}}.$ Using Eq.~\ref{eq: local gradient}, together with $\hat{\mathbf{P}}_\phi = \mathbf{P}_\Gamma + \mathcal{O}(\epsilon^2)$ and $\vn \cdot \hat{\mathbf{P}}_\phi = \mathcal{O}(\epsilon^2)$, yields
\begin{equation*}
    \hat{\mathbf{P}}_\phi \mathbin{:} \nabla \hat{\mathbf{v}} = \nabla_\Gamma \cdot (\mathbf{P}_\Gamma \hat{\mathbf{v}}) - (\nabla_\Gamma \cdot \vP_\Gamma)\cdot\hat{\mathbf{v}} + \mathcal{O}(\epsilon).
\end{equation*}
Plugging in the expansion of $\hat{\mathbf{v}}$ and using $\nabla_\Gamma \cdot \vP_\Gamma = -\kappa \vn$, 
with curvature $\kappa$, we obtain 
\begin{align*}
    \hat{\mathbf{P}}_\phi \mathbin{:} \nabla \hat{\mathbf{v}} &= \nabla_\Gamma \cdot \left(\vP_\Gamma \hat{\vv}^{(0)}\right) + \kappa \vn \cdot \hat{\vv}^{(0)} + \mathcal{O}(\epsilon) \\ 
    &= \nabla_\Gamma \cdot \hat{\vv}^{(0)} + \mathcal{O}(\epsilon).
\end{align*}
Hence, the diffuse stress tensor reduces directly to its sharp counterpart evaluated at the leading-order velocity:
\begin{equation}
    \hat{\mathbf{S}}_\phi = \mathbf{S}_\Gamma(\hat{\mathbf{v}}^{(0)}) + \mathcal{O}(\epsilon).
\end{equation}
Note, $\mathbf{S}_\Gamma$ is in the sharp interface limit a purely tangential tensor, satisfying $\mathbf{S}_\Gamma \cdot \vn = \mathbf{0}$. In the diffuse interface setting $\hat{\mathbf{S}}_\phi$ is hence tangential to the highest order $\hat{\mathbf{S}}_\phi \cdot \vn = \mathcal{O}(\epsilon) $. 

Next we evaluate the contribution of the viscous stress $\hat{\mathbf{S}}_\phi$ to the momentum balance. It holds
\begin{align}
    &\nabla \cdot \left( \vert\nabla \phi \vert \hat{\mathbf{S}}_\phi \right) \nonumber\\ =& \dfrac{1}{\epsilon} \vn \cdot \partial_z \left( \dfrac{1}{\epsilon} \partial_z \hat{\phi}^{(0)} \hat{\mathbf{S}}_\phi\right) + \nabla_\Gamma \cdot \left( \dfrac{1}{\epsilon} \partial_z \hat{\phi}^{(0)} \hat{\mathbf{S}}_\phi\right) \nonumber\\
    =& \dfrac{1}{\epsilon^2} \left(\partial_{zz} \hat{\phi}^{(0)} \hat{\mathbf{S}}_\phi \cdot \vn + \partial_z \hat{\phi}^{(0)}\vn \cdot \partial_z \hat{\mathbf{S}}_\phi  \right) \nonumber \\ &+ \dfrac{1}{\epsilon} \nabla_\Gamma \cdot \left( \partial_z \hat{\phi}^{(0)} \hat{\mathbf{S}}_\phi \right).\label{eq: viscous force in local cs}
\end{align}
Since $\vn_\phi = \vn + \mathcal{O}(\epsilon^2)$ holds, $\hat{\mathbf{S}}_\phi$ is tangential at least to the highest two orders, such that the first bracket on the right hand side of Eq.~\eqref{eq: viscous force in local cs} is of order $\mathcal{O}(\epsilon^2)$. Hence, the contribution of the first term is of order $\mathcal{O}(1)$ and does not enter the leading-order $\mathcal{O}(1/\epsilon)$ balance. We obtain to leading order $\frac{1}{\epsilon}$
\begin{equation}\label{eq: viscous force leading order}
    \nabla_\Gamma \cdot \left( \partial_z \hat{\phi}^{(0)} \hat{\mathbf{S}}_\phi \right).
\end{equation} Since $\nabla_\Gamma \partial_z \hat{\phi}^{(0)} = \partial_z \nabla_\Gamma \hat{\phi}^{(0)} = 0$, integration of \eqref{eq: viscous force leading order} over $z$ gives
\begin{equation*}
    \nabla_\Gamma \cdot \hat{\vS}^{(0)} \int_{-\infty}^{\infty} \partial_z \hat{\phi}^{(0)} \, \text{d}z,
\end{equation*} where we have used that $\hat{\vS}^{(0)}$ is independent of $z$, since both $\vv$ (see Sec.~\ref{sec: asymptotic ana v}) and $\vn$ are independent of $z$ to highest order. From Eq.~\eqref{eq: zero order phi hat} we know that $\int_{-\infty}^{\infty} \partial_z \hat{\phi}^{(0)} \, \text{d}z = 1$, hence we recover at order $\frac{1}{\epsilon}$ the contribution $\nabla_\Gamma \cdot \hat{\vS}^{(0)}$. 

The remaining terms of the momentum balance at this order are derived in \cite[see Eq. 4.30]{AGG12}. Putting everything together gives 
\begin{align*}
    &\left[ -\eta \left( \nabla \vv^{(0)} + (\nabla \vv^{(0)})^T \right) + \mathbf{I} p^{(0)} \right] \cdot \vn \\
    =& \sigma \kappa \vn + \nabla_\Gamma \cdot \hat{\vS}^{(0)}
\end{align*}
recovering the desired jump condition from Eq.~\eqref{eq: sharp interface condition}.

\subsubsection{Continuity of velocity} \label{sec: asymptotic ana v}

Finally we show that the phase-field model preserves the continuity of the velocity $\vv$ across the interface.

Applying Eq.~\eqref{eq: local gradient} to the incompressibility condition $\nabla \cdot \vv$ in the inner region, we obtain
\begin{equation*}
    \dfrac{1}{\epsilon} \partial_z (\hat{\vv} \cdot \vn) + \nabla_\Gamma \cdot \hat{\vv} = 0.
\end{equation*} This leads to the order-by-order relations:
\begin{align*}
    \partial_z (\hat{\vv}^{(0)} \cdot \vn) &= 0 \qquad \text{at } \mathcal{O}(1/\epsilon), \\
    \hat{\vv}^{(1)}_z \cdot \vn + \nabla_\Gamma \cdot \hat{\vv}^{(0)} &= 0 \qquad\text{at } \mathcal{O}(1),
\end{align*} and continuing similarly for higher orders. To finalize the proof, we need to show that the tangential components of the velocity, namely $\vP_\Gamma \hat{\vv}^{(0)}$, are also independent of $z$. This follows from the viscous term in the Navier--Stokes equations. As demonstrated in \cite{AGG12}, the viscous term dominates the inner expansion at the highest order. Using that the contribution of the surface stress to the momentum is of lower order (see Eq.~\eqref{eq: viscous force leading order}), at the leading order, $\mathcal{O}(1/\epsilon^2)$, the Navier--Stokes equations become
\begin{equation*}
    \partial_z \left(\eta \hat{\vv}^{(0)}_z \cdot \vn\right) \vn + \partial_z \left( \eta \partial_z \left( \vP_\Gamma \hat{\vv}^{(0)} \right)\right) = 0.
\end{equation*} Given that the first term is zero, it follows that $\eta \partial_z\left( \vP_\Gamma \hat{\vv}^{(0)} \right) = \text{constant.} $ Taking $z \rightarrow \pm \infty$ and using that the matching condition Eq.~\eqref{eq: matching cond velocities} requires $\vP_\Gamma \hat{\vv}^{(0)}$ to be bounded, we find that the constant is equal to zero.  This concludes the proof.

\section{Numerical method}\label{sec:discretization}
\subsection{Energy-stable time discretization}
For the time discretization we chose an equidistant time partitioning with time step size $\Delta t$. For some time-dependent scalar variable $f$ the time derivative can then be approximated by the discrete time derivative $d_t f^n \coloneqq ({f^n -f^{n-1}})/{\Delta t}  \approx  \partial_t f$ at the $n$-th time step.

As shown in Sec~\ref{sec:energy stability model}, the presented phase-field model for viscous surfaces is energy stable. 
To achieve unconditional energy stability at the discrete level, we employ a discrete gradient approach. We modify the discretization of the chemical potential by replacing $W^\prime(\phi)$ with a non-linear difference quotient that exactly captures the energy difference between two consecutive time steps.
Thus, we construct a discrete function $\chi(\phi^n, \phi^{n-1})$ that satisfies
\begin{equation} \label{eq:exact_diff}
    W(\phi^n) - W(\phi^{n-1}) = \chi(\phi^n, \phi^{n-1}) (\phi^n - \phi^{n-1}).
\end{equation}
By expanding the difference of the potentials as a difference of squares, we obtain 
a first order approximation of $W'(\phi^n)$ as:
\begin{align} 
    \chi(\phi^n, \phi^{n-1}) \coloneqq & 18 (\phi^n + \phi^{n-1} - 1) \nonumber \\ &\cdot \left( (\phi^n)^2 - \phi^n + (\phi^{n-1})^2 - \phi^{n-1} \right). \label{eq:chi_def}
\end{align}
In the following, we present a stable, unconditionally energy dissipative, semi-implicit Euler scheme for the Navier--Stokes--Cahn--Hilliard  Eqs.~\eqref{eq: NS balance of momentum} to \eqref{eq:diffuse_stress_tensor} with surface viscosity.
For simplicity, we present the method for constant densities, $\rho(\phi) = \rho = const$.
The extension to variable densities can be carried out by a standard splitting of the convective term, analogously as in \cite{grun2014two}.
\newpage
The time-discrete system reads
\begin{strip}
\begin{align}
    d_t\phi^{n} 
    =& \, \nabla\cdot\bigl(M\nabla\mu^n\bigr)
       - \nabla\phi^{n-1}\cdot\mathbf v^{n}, 
       \label{eq:time discr. NSCH 1} \\[4pt]
    \mu^n 
    =& -\epsilon\Delta\phi^n + \epsilon^{-1}\chi(\phi^n,\phi^{n-1}), 
       \label{eq:time discr. NSCH 2} \\[4pt]
    \nabla\cdot\mathbf v^{n} 
    =& \, 0,   
       \label{eq:time discr. NSCH 3} \\[4pt]
    \rho\bigl(d_t\mathbf v^n + \mathbf v^{n-1}\!\cdot\!\nabla\mathbf v^n\bigr)
    =& -\nabla p^n +\nabla\cdot\bigl(2\eta(\phi^{n-1})\mathbf D(\mathbf v^n)\bigr) \label{eq:time discr. NSCH 4}  \\
    &+ \sigma\,\mu^n\nabla\phi^{n-1} +\nabla \cdot (\vert \nabla \phi ^{n-1}\vert \vS_{\phi}^n), \nonumber
\end{align} 
\end{strip}
 where $\mathbf{D}(\vv^n) =\frac{1}{2}\left(\nabla\textbf{v}^n  +(\nabla\textbf{v}^n)^T\right)$ denotes the rate of deformation tensor,
 \begin{align}
    \vS_{\phi}^n =& (\eta_b - \eta_s) (\mathbf{P}_\phi^{n-1} \mathbin{:} \nabla \mathbf{v}^n) \mathbf{P}_\phi^{n-1} \nonumber \\
    &+  \eta_s \vP_\phi^{n-1} \left( \nabla \textbf{v}^n +(\nabla \textbf{v}^n)^T \right) \vP_\phi^{n-1}
 \end{align} is the time-discrete viscous surface stress,
 and $\mathbf{P}_\phi^{n-1} = \mathbf{I} - \vn_\phi^{n-1} \otimes \vn_\phi^{n-1}$ is the time-discrete projection tensor.
 As shown in \cite{ALAND2014_TimeIntegration} solving this fully-coupled system monolithically has a strong stabilizing effect, allowing larger time steps with less nonlinear subiterations.

Next, we prove the energy stability of the time-discrete Navier--Stokes--Cahn--Hilliard system \eqref{eq:time discr. NSCH 1} to \eqref{eq:time discr. NSCH 4}. Let $\left \langle \cdot,\cdot \right\rangle_{L_2}$ denote the $L_2$-inner product on $\Omega$. We assume appropriate boundary conditions (e.g., periodic or no-slip/no-flux) such that boundary integrals vanish during integration by parts.

First, we test Eq.~\eqref{eq:time discr. NSCH 1} with $\sigma \mu^n$. Using the homogeneous Neumann boundary condition ${\nabla \mu^n \cdot \vn = 0}$, we obtain:
\begin{align}
    \sigma \left \langle \frac{\phi^n - \phi^{n-1}}{\Delta t}, \mu^n \right\rangle_{L_2} =& -\sigma M \| \nabla \mu^n \|^2_{L_2} \nonumber\\ &- \sigma \left\langle \textbf{v}^n \cdot \nabla \phi^{n-1}, \mu^n \right\rangle_{L_2}. \label{eq:test1}
\end{align}

Next, we test Eq.~\eqref{eq:time discr. NSCH 2} with $\sigma \frac{\phi^n - \phi^{n-1}}{\Delta t}$. Using integration by parts for the Laplacian term, the exact identity from \eqref{eq:exact_diff}, and the polarization identity $(a - b) \cdot a = \frac{1}{2} \left( |a|^2 - |b|^2 + |a - b|^2 \right)$, we obtain:
\begin{align}
    &\sigma \left \langle \mu^n, \frac{\phi^n - \phi^{n-1}}{\Delta t} \right\rangle_{L_2} \nonumber \\ =& \sigma \epsilon \left \langle \nabla \phi^n, \frac{\nabla \phi^n - \nabla \phi^{n-1}}{\Delta t} \right\rangle_{L_2} \nonumber \\
    &+ \frac{\sigma}{\epsilon \Delta t} \int_\Omega \chi(\phi^n, \phi^{n-1}) (\phi^n - \phi^{n-1}) d\vx \nonumber \\
    =& \frac{\sigma \epsilon}{2 \Delta t} \int_\Omega \vert \nabla \phi^n \vert^2 - \vert \nabla \phi^{n-1} \vert^2 d\vx \nonumber \\ &+ \frac{\sigma \epsilon}{2 \Delta t} \int_\Omega\vert \nabla \phi^n - \nabla \phi^{n-1}\vert^2 d\vx \nonumber \\
    &+  \frac{\sigma}{\epsilon \Delta t} \int_\Omega \left( W(\phi^n) - W(\phi^{n-1}) \right) d\vx. \label{eq:test2}
\end{align}
Equating \eqref{eq:test1} and \eqref{eq:test2} and rearranging the terms yields the discrete energy balance of the Cahn--Hilliard subsystem:
\begin{align}
    &\frac{\sigma \epsilon}{2 \Delta t} \int_\Omega \left( | \nabla \phi^n |^2 - | \nabla \phi^{n-1} |^2 \right) \, d\vx \nonumber \\&+ \frac{\sigma}{\epsilon \Delta t} \int_\Omega \left( W(\phi^n) - W(\phi^{n-1}) \right) \, d\vx \nonumber \\
    =& - \sigma M \| \nabla \mu^n \|^2_{L_2} - \frac{\sigma \epsilon}{2 \Delta t} \| \nabla \phi^n - \nabla \phi^{n-1} \|^2_{L_2} \nonumber \\ &- \sigma \left\langle \textbf{v}^n \cdot \nabla \phi^{n-1}, \mu^n \right\rangle_{L_2}. \label{eq:CH_energy_balance}
\end{align}
For the balance of momentum Eq.~\eqref{eq:time discr. NSCH 4}, we use $\textbf{v}^n$ as a test function. The pressure term vanishes due to the incompressibility constraint \eqref{eq:time discr. NSCH 3}. The advection term vanishes for appropriate boundary conditions as well, since with the incompressibility $\nabla \cdot \vv^{n-1} = 0$ 
it holds
\begin{align*}
    &\int_\Omega \rho  (\vv^{n-1} \cdot \nabla)\mathbf{v}^{n} \cdot \vv ^{n} \, d\vx \\=& \int_\Omega \frac{1}{2} \rho \vv^{n-1} \cdot \nabla(\vert\mathbf{v}^{n}\vert^2) \, d\vx \\
    =& 0
\end{align*}
 Applying again the polarization identity and the same ideas as in Sec.~\ref{subsec: kinetic energy dissipation continuous case} (cf. Eqs.~\eqref{eq: S times nabla v} and \eqref{eq: S times nabla v after splitting}) for Eq.~\eqref{eq:time discr. NSCH 4} gives:
\begin{align}
    & \quad \frac{1}{2 \Delta t} \int_\Omega \rho \left( |\textbf{v}^n|^2 - |\textbf{v}^{n-1}|^2 + |\textbf{v}^n - \textbf{v}^{n-1}|^2 \right) d\vx \nonumber \\
    =&-  \int_\Omega 2 \eta(\phi^{n-1}) \| \mathbf{D}(\vv^n) \|^2_F d\vx \nonumber \\ &+ \sigma \left \langle \mu^n \nabla \phi^{n-1}, \textbf{v}^n \right\rangle_{L_2} \nonumber \\ &- \int_\Omega \vert\nabla \phi^{n-1}\vert  \Big( \left(\eta_b + (3-d) \eta_s \right)(\nabla_{\Gamma, \phi^{n-1}} \cdot \mathbf{v}^{n})^2  \nonumber \\ 
    &\qquad + 2\eta_s \|\mathbf{\bar{\vD}}_\phi^{n}\|_F^2 \Big) d\vx  \label{eq:test3}
\end{align}
Now, we sum the Cahn--Hilliard energy balance \eqref{eq:CH_energy_balance} and the momentum balance \eqref{eq:test3}. Here, the advective coupling terms $\pm\sigma \left \langle \mu^n \nabla \phi^{n-1}, \textbf{v}^n \right\rangle$ cancel out exactly. 

We define the total discrete free energy at time step $n$ as:
\begin{equation}
    E_{tot}^n \coloneqq \int_\Omega \left( \frac{\rho}{2} |\textbf{v}^n|^2 + \frac{\sigma \epsilon}{2} |\nabla \phi^n|^2 + \frac{\sigma}{\epsilon} W(\phi^n) \right) d\vx.
\end{equation}
The resulting discrete energy law simplifies to:
\begin{align} 
    &\frac{E_{tot}^n - E_{tot}^{n-1}}{\Delta t} \nonumber \\=& - \sigma M \| \nabla \mu^n \|^2_{L_2} -  \int_\Omega 2 \eta(\phi^{n-1}) \| \vD(\vv^n) \|^2_F \, d\vx \nonumber \nonumber \\ &- \int_\Omega|\nabla \phi^{n-1}| \Big(  \left(\eta_b + (3-d) \eta_s \right)(\nabla_{\Gamma, \phi^{n-1}} \cdot \mathbf{v}^{n})^2 \nonumber \\& \qquad + 2\eta_s \|\mathbf{\bar{\vD}}_\phi^{n}\|_F^2 \Big) \, d\vx \nonumber \\ &- \mathcal{D} \nonumber \\ \le& \ 0, \label{eq:final_energy_law}
\end{align}
where the term 
\begin{align*}
    \mathcal{D} =& \frac{1}{2\Delta t} \int_\Omega \rho |\textbf{v}^n - \textbf{v}^{n-1}|^2 d\vx \\ &+ \frac{\sigma \epsilon}{2 \Delta t} \| \nabla \phi^n - \nabla \phi^{n-1} \|^2_{L_2} \\ \ge & 0
\end{align*}
represents the non-negative numerical dissipation of order $\mathcal{O}(\Delta t)$ introduced by the first-order backward Euler time integration. Because the right-hand side of \eqref{eq:final_energy_law} is strictly non-positive, the sequence of total energies $E_{tot}^n$ is monotonically decreasing, proving that the proposed fully-coupled non-linear scheme is thermodynamically consistent. 

\subsection{Finite element discretization}
In space we discretize the presented models using a finite element approach. Let $T_h$ be a triangulation of the domain $\Omega$ and $P_m(T)$, $m \in \mathbb{N}$, denote the space of polynomials of total degree at most $m$ on a triangle $T \in T_h$. 
We use an adaptive refinement strategy for the mesh around the interface, where coarsening and refinement of the mesh take place such that the region $\phi \in \left[0.01, 0.99 \right]$  across the interface is resolved by at least five degrees of freedom. Away from the interface the grid size is fixed to a coarser value $\Delta \vx$. 

We use Lagrange-$P_1$ finite elements for the pressure $p$ and $P_2$ elements for the other variables. The corresponding finite element spaces are
\begin{align*}
    Q_h &= \lbrace q \in L_0^2(\Omega) \ \vert \ q\bigr|_T \in P_1(T)\rbrace, \\
    U_h &= \lbrace u \in C^0(\bar{\Omega})\cap H_0^1(\Omega) \ \vert \ u\bigr|_T \in P_2(T)\rbrace, \\
    Y_h &= \lbrace \psi \in C^0(\bar{\Omega}) \ \vert \ \psi\bigr|_T   \in P_2(T)\rbrace.
\end{align*}
The weak formulations for  the Eqs.~\eqref{eq:time discr. NSCH 1} to \eqref{eq:time discr. NSCH 4} including the viscous surface stress term $\nabla \cdot (\vert \nabla \phi ^{n-1}\vert \vS_{\phi}^n)$  read as follows. 
\\[\baselineskip]
\textbf{Navier--Stokes--Cahn--Hilliard system:} Find $\left(\phi^n,\mu^n, p^n, \textbf{v}^n\right) \in Y_h \times Y_h \times Q_h \times (U_h)^2 $ such that for all $\left(\psi,\xi, q, \textbf{u}\right) \in Y_h \times Y_h \times Q_h \times (U_h)^2 $ 
\begin{strip}  
\begin{align*}
    \left\langle d_t \phi^n, \psi \right\rangle_{L_2} =\,& \left\langle  -M \nabla \mu^n, \nabla \psi \right\rangle_{L_2} - \left\langle\nabla \phi^{n-1} \cdot \textbf{v}^{n}, \psi\right\rangle_{L_2}, \\ 
    \left\langle\mu^n,\xi\right\rangle_{L_2} - \left\langle \epsilon \nabla \phi^n, \nabla \xi \right\rangle_{L_2} =\,& \epsilon^{-1} \left\langle\chi(\phi^n,\phi^{n-1}), \xi \right\rangle_{L_2}, \\ 
    \left\langle\nabla \cdot\textbf{v}^{n},q \right\rangle_{L_2} =\,& 0, \\   
    \begin{split}
        \rho \left( \left\langle d_t \textbf{v}^n, \textbf{u} \right\rangle_{L_2} + \left\langle (\textbf{v}^{n-1} \cdot \nabla) \textbf{v}^n, \textbf{u} \right\rangle_{L_2} \right)
        =\,& - \left\langle 2 \eta(\phi^{n-1}) \mathbf{D}(\mathbf{v}^n), \nabla \textbf{u} \right\rangle_{L_2}  + \left\langle p^n, \nabla \cdot \textbf{u} \right\rangle_{L_2} \\ &+ \left\langle \sigma \mu^n \nabla \phi^{n-1} 
        , \textbf{u}\right\rangle_{L_2} \\ &+ (\eta_b - \eta_s) \left\langle |\nabla \phi^{n-1}| (\mathbf{P}_{\phi}^{n-1} : \nabla \mathbf{v}^n), (\mathbf{P}_{\phi}^{n-1} : \nabla \mathbf{u}) \right\rangle_{L_2} \\&+ 2 \eta_s \left\langle |\nabla \phi^{n-1}| \mathbf{P}_{\phi}^{n-1} \mathbf{D}(\mathbf{v}^n) \mathbf{P}_{\phi}^{n-1}, \nabla \mathbf{u} \right\rangle_{L_2}.
    \end{split}
\end{align*}
\end{strip}
We solve the resulting nonlinear system of equations using a Newton method as described in the following section.

\subsection{Nonlinear solution procedure}

The finite element discretization of the Navier--Stokes--Cahn--Hilliard system results in a nonlinear algebraic system
\begin{equation*}
    F(U)=0,
\end{equation*}
where the solution vector $U=(\phi,\mu,p,\vv)^T$
contains all discrete unknowns at the current time step.

This system is solved by a standard Newton iteration. In every Newton step $k$, the Jacobian of the current nonlinear residual is assembled and the linear correction
\begin{equation*}
    J(U^k)\,\delta U = -F(U^k)
\end{equation*}
is computed, where $J(U^k)$ denotes the exact Jacobian matrix evaluated at the current iterate $U^k$. The solution is subsequently updated according to
\begin{equation*}
    U^{k+1}=U^k+\delta U .
\end{equation*}
Due to the semi-implicit time discretization introduced in Eqs.~\eqref{eq:time discr. NSCH 1} to \eqref{eq:time discr. NSCH 4}, the strongest nonlinear couplings between the fluid and the phase-field are linearized.
Most notably, the geometric components of the viscous surface stress tensor -- such as the projection operators $\vP_\phi^{n-1}$ and the surface delta function $\vert \nabla \phi^{n-1} \vert$ -- are evaluated explicitly at the previous time step. Consequently, the otherwise highly complex surface stress $\nabla \cdot (\vert \nabla \phi ^{n-1}\vert \vS_{\phi}^n)$ becomes completely linear with respect to the velocity $\mathbf{v}^n$. 
Similarly, the phase-dependent viscosity and the convective term are linearized. 
For the Newton solver, these terms merely act as fixed, spatially varying coefficients, meaning only the remaining implicit nonlinear term $\chi(\phi^n,\phi^{n-1})$ needs to be differentiated during the Jacobian assembly. 

The Newton iteration is terminated once the relative correction satisfies 
\begin{equation*}
    \frac{\Vert\delta U\Vert}{\Vert U\Vert+1} < {\rm tol},
\end{equation*}
or after a prescribed maximum number of iterations $N_{max}$. If not stated otherwise the relative tolerance is set to $\text{tol} = 10^{-8}$ and $N_{max} = 15$ throughout this paper. Because the semi-implicit treatment successfully decouples the severe geometric nonlinearities, the solver is highly robust and the relative stopping criterion is in practice typically reached within only a few iterations.

The resulting linear systems are solved using the sparse direct solver UMFPACK~\cite{Davis_UMFPACK}. Note, to ensure regularity of the linear systems a small constant was added to the numerical approximation $\vert \nabla \phi \vert$ of the surface delta function. All discretizations and nonlinear solvers are implemented within the finite element toolbox DUNE/AMDiS~\cite{PraetoriusAMDiS,AMDIS_2007,Witkowski_adaptive_parallel_FEM}.

\section{Numerical results}\label{sec:Numerical results}

\subsection{Toy example: A Viscous Surface in Shear Flow} \label{sec: viscous band}
As a first experiment to illustrate the influence of the viscous surface force, we simulate the deformation of a simple one-dimensional interface curve due to a surrounding shear flow. 

\begin{figure*}
\centering
\includegraphics[width=\linewidth ]{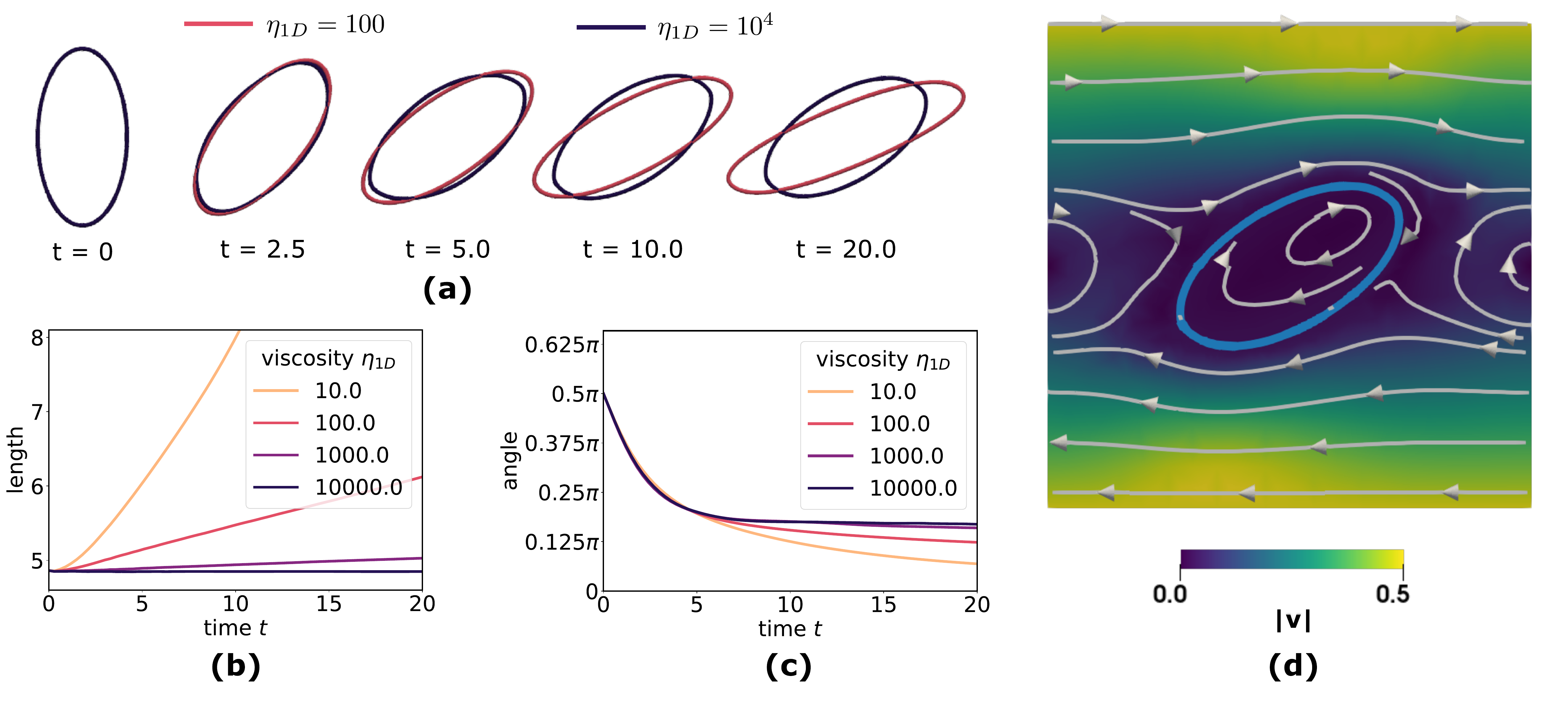}
\caption{Dynamic response of a 1D viscous interface in an ambient shear flow. \textbf{(a)} Snapshots of the interface deformation at different time steps for two different values of the surface dilational viscosity $\eta_{\rm 1D}=\eta_b+\eta_s$. \textbf{(b)} Evolution of the interface length over time for varying $\eta_{\rm 1D}$. High viscosities ($\eta_{\rm 1D} \ge 10^4$) enforce inextensibility. \textbf{(c)} Evolution of the angle between the major ellipse axis and the flow direction ($x$-axis). Highly viscous interfaces reach a steady-state angle that balances hydrodynamic torque and internal resistance against stretching. \textbf{(d)} Snapshot of the highly viscous interface ($\eta_{\rm 1D} = 10^4$) in its steady state with the surrounding flow colored by the velocity magnitude $\vert \vv \vert$. 
\textbf{Simulation parameters:} $\eta_0 = \eta_1 = 1.56$, $\rho_0 = \rho_1 = 1.0$, $\epsilon = 0.03$, $M = 10^{-5}$, $\sigma = 0$, $\Delta \vx = 0.4$, $\Delta t = 0.001$, $N_{max} = 1$}
\label{fig:viscous_band}
\end{figure*}

The initial interface of the phase-field $\phi$ is prescribed by an elliptic shape (axes with lengths of $1$ and $0.5$) centered in the domain $\Omega_{\text{2D}} = \left[-2,2 \right]^2$, see Fig.~\ref{fig:viscous_band}a (t=0).  
To drive the system, we impose an external stationary shear flow with a shear rate of $\dot{\gamma} = 0.5$. Numerically, this ambient flow is initialized across the domain and maintained on the domain walls ($\delta \mathbf{v} = 0$ as boundary condition of the Newton iteration).

As mentioned in Sec.~\ref{sec:sharp interface model} the rheology of the 1-dimensional interface is governed by a single parameter $\eta_{\rm 1D} = \eta_b+\eta_s$.
To illustrate the effect of the surface viscosity we choose dimensionless viscosities from $\eta_{\rm 1D} \in \{10^i : i = 1,2,3,4\}$. 
Fig.~\ref{fig:viscous_band} illustrates the numerical results, showing how the viscous band rotates and deforms due to the imposed shear flow.  
For smaller $\eta_{\rm 1D}$ the surface gets stretched rapidly, indicated by a strong change of interface length (Fig.~\ref{fig:viscous_band}b). With increasing $\eta_{\rm 1D}$ the shear force is eventually balanced by the viscous surface force, s.~t. the surface tends towards a state of inextensibility. For $\eta_{\rm 1D} = 10^4$ the surface does not remarkably change in length anymore.

Besides the length change, we track the change of the angle between the major axis of the ellipse and the $x$-axis. 
As depicted in Fig.~\ref{fig:viscous_band}c, the major axis is initially oriented vertically at an angle of $0.5\pi$. Driven by the shear flow, the interface immediately starts rotating towards the direction of flow. In the initial phase ($t \lesssim 5$), the rotational trajectory is almost independent of the surface viscosity. 
For larger times ($t > 5$), however, the surface viscosity exerts a significant influence on the alignment dynamics. For low values of $\eta_{\rm 1D}$ (e.g., $\eta_{\rm 1D} \le  10$), the surface continues to rotate and aligns almost horizontally with the flow axis ($\text{angle} \to 0$) as it experiences severe stretching. Conversely, with increasing surface viscosity ($\eta_{\rm 1D} \ge 10^3$), the resistance against dilational deformation counteracts the shear torque. Consequently, the rotation decelerates and the inclination angle approaches a steady-state plateau around $0.16\pi$. 

This equilibrium orientation as well as the constant surface length confirm that high surface viscosities enforce inextensibility, balancing the external hydrodynamic forces and preventing further deformation as well as complete alignment with the shear direction.
The corresponding flow profile shows a tumbling motion (Fig.~\ref{fig:viscous_band}d).

\subsection{Validation of Surface Viscous Flows on a Rigid Sphere} \label{sec:validation rigid sphere}

To quantitatively validate the presented viscous surface model, we consider the canonical problem of a flow on a stationary, rigid sphere. The numerical results are benchmarked against exact analytical solutions for active fluid surfaces in viscous environments, as derived by Mietke \cite[chapter 3]{mietke2018dynamics} and in \cite[SI chapter 3.3]{mietke_min_model19}. 

\subsubsection{Analytical Solution of Surface/Bulk Viscous Flow}
The analytical results were derived for Stokes flow of a spherical viscous surface embedded in an infinitely large fluid domain. Moreover, the analytical model assumes a vanishing normal velocity, $\vv\cdot\mathbf{n}=0$ on $\Gamma$ (cf. \cite[Eq. (3.8)]{mietke2018dynamics}), with flow induced by a stationary tangential driving force along the surface, given by
\begin{align}
    \mathbf{f}_{\rm drive} = \nabla_\Gamma \mathcal{P}_l(\cos\theta),
\end{align}
where $\mathcal{P}_l$ is the Legendre polynomial of degree $l$ and $\theta$ is the polar angle along the spherical surface. 
For a spherical surface of radius $R=1$, the derived analytical solution predicts an axisymmetric solution for tangential surface velocity (in the $\theta$-direction), given by 
\begin{strip}
   \begin{equation} \label{eq: analytical tangential velocity}
    v_\theta = \dfrac{\partial_\theta \mathcal{P}_l(\cos\theta)}{(\eta_0+\eta_1) (1+2l)-\eta_0  \delta_{1,l}+(\eta_b+\eta_s)l(l+1) - 2\eta_s},
\end{equation} 
\end{strip} cf. \cite[Eq. (3.19) with $R=\xi=1$]{mietke2018dynamics}.
The solution depends on the chosen mode $l$, the surface and bulk viscosities. 
Note that for $l\neq 1$ the Kronecker delta in Eq.~\eqref{eq: analytical tangential velocity} is $\delta_{1,l}=0$, such that the bulk viscous contribution only depends on the sum $\eta_0+\eta_1$. 
In our numerical tests, we choose $\eta_0\ll\eta_1$ to eliminate the influence of the outer domain boundary (which is infinitely far in the analytical benchmark). 

\begin{figure*}
    \centering
    \includegraphics[width=\linewidth]{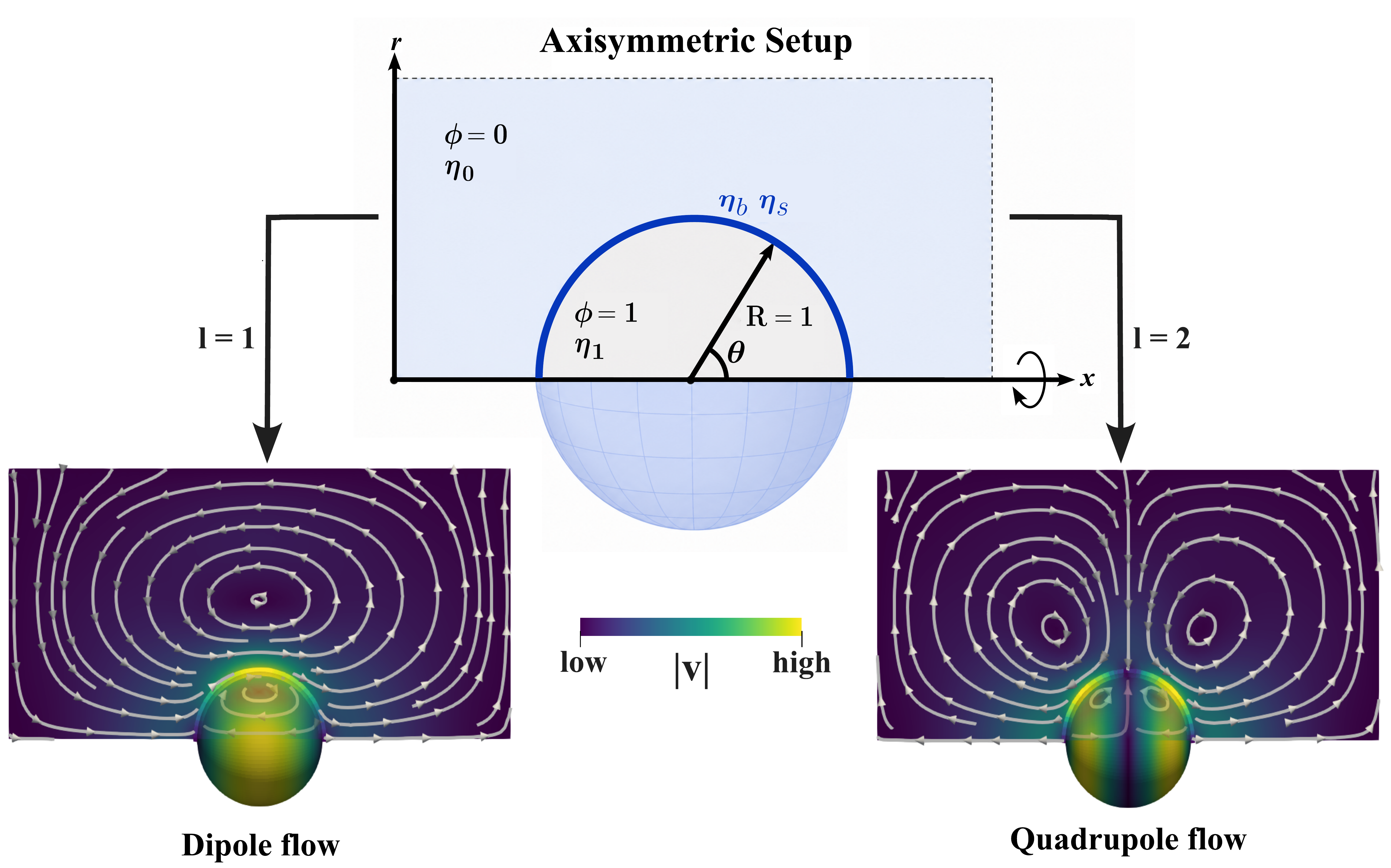}
    \caption{Axisymmetric benchmark setup of surface-tension-driven flow around a stationary, rigid sphere. \textbf{Top:} Sketch of the general setup showing the two-dimensional meridian plane. The diffuse interface (blue semicircle) represents the spherical viscous surface with viscosities $\eta_b, \eta_s$. 
    \textbf{Bottom:} Flow is driven by the force $\mathbf{f}_{\rm drive}= \nabla_\Gamma   \mathcal{P}_l(\cos\theta)$ with $l=1$ (left) and $l=2$ (right), leading to dipole flow and quadrupole flow, respectively}
    \label{fig:axisym setup}
\end{figure*}

In our benchmark studies, we focus on the fundamental modes $l=1$ and $l=2$, which enable us to probe individual contributions of surface dilation and surface shear viscosity.
Fig.~\ref{fig:axisym setup} illustrates the resulting axisymmetric flows. 
For $l = 1$, the first Legendre polynomial $\mathcal{P}_1(\cos \theta) = \cos \theta$ establishes a unidirectional force from one pole to the other (here left to right) driving a dipole flow field in the bulk phase. 
Conversely, for $l = 2$, the profile follows $\mathcal{P}_2(\cos \theta) = \frac{1}{2}(3\cos^2\theta - 1)$. This symmetric force drives the fluid from the equator (located at $x = 0$) towards both poles, generating two distinct, counter-rotating vortexes.

\subsubsection{Diffuse-interface Formulation of the Benchmark Problem}
To construct a diffuse-interface formulaton of the benchmark problem, we impose the driving force as 
$\mathbf{f}_{\text{drive}} = |\nabla \phi| \vP_\phi\nabla \mathcal{P}_l(\cos\theta)$, which is added to the right hand side of the balance of momentum Eq.~\eqref{eq: NS balance of momentum}.
To include the vanishing normal velocity in the diffuse-interface framework, we augment our model with the additional force term
\begin{equation}\label{eq:penalty term}
    \lambda |\nabla\phi|
    (\mathbf v\cdot\mathbf n_\phi)
    \mathbf n_\phi
\end{equation}
on the left-hand side of the momentum equation \eqref{eq: NS balance of momentum}, where $\lambda \gg 1$ is a penalty parameter. This term penalizes velocity components normal to the diffuse interface and therefore approximately enforces the impermeability condition from the benchmark case:
\begin{equation*}
    \mathbf v\cdot\mathbf n_\phi \approx 0.
\end{equation*}

As the benchmark is in the Stokes regime with fixed geometry, the problem is entirely time-independent. Consequently, we do not solve the Cahn--Hilliard system; instead, the diffuse interface $\phi$ is analytically prescribed as a static background field representing a circle of unit radius ($R=1$) in the two-dimensional axisymmetric meridian plane. The corresponding computational domain for the numerical studies is $\Omega_{\text{2D}} = \left[-2, 2 \right]\times \left[ 0, 2\right]$.

\subsubsection{Measure of deviation}
To measure the approximation error in our numerical results ($v_{\theta,\phi} = |\mathbf{P}_\phi\vv|$) compared to the analytical reference solutions ($v_\theta$), we use the relative $L_2$ error:
\begin{equation*}
    \text{err}_{tot} = \dfrac{\Vert v_\theta - v_{\theta,\phi}\Vert_{\delta}}{\Vert v_\theta \Vert_{\delta}},
\end{equation*} where $\Vert{} u \Vert{}_{\delta} = \sqrt{\langle u, u \rangle_{\delta}}$ denotes an interface-weighted $L_2$-norm with $\langle u, w \rangle_{\delta} = \int_{\Omega} \vert\nabla\phi\vert \, u \, w \,  r \, \mathrm{d}x$.
Because the analytical solution $v_\theta$ is originally defined strictly on the sharp interface, we implicitly extend it constantly in the normal direction across the diffuse interface band to evaluate this volume integral.
The norm of the numerical solution now reads $\Vert v_{\theta,\phi} \Vert_{\delta} = \left( \int_\Omega \vert \nabla\phi\vert \vert \mathbf{P}_\phi \vv  \vert^2 r \text{d}x \right)^{1/2}$. Note that for the phase-field representation employed here, $\vert \nabla \phi \vert$ acts as a diffuse approximation of the surface delta distribution, $\delta \approx \vert\nabla \phi \vert$.

\subsubsection{Choosing the Penalty Parameter}
We conducted a preliminary parameter study to determine a suitable value for the penalty parameter $\lambda$, to effectively suppress normal velocities at the interface, see Fig.~\ref{fig:validation1}a. 
Setting $\lambda=0$ leads to a severe violation of this condition, yielding a high normal-to-tangential velocity ratio
\begin{equation*}
    \mathcal{R}_{n/t}  = \frac{\int_{\Omega} \vert{}\mathbf{v} \cdot \mathbf{n}_\phi\vert{}^2 \, \mathrm{d}x}{\int_{\Omega} \vert{}\mathbf{v}_\theta\vert{}^2 \, \mathrm{d}x} \approx 1
\end{equation*} and a high relative approximation error $\text{err}_{tot} = \Vert v_\theta - v_{\theta,\phi}\Vert_{\delta}/\Vert v_\theta \Vert_{\delta} \approx 0.365$. 

For $\lambda \in [200, 400]$, the normal velocity contribution drops well below $0.1\%$ ($\mathcal{R}_{n/t} < 10^{-3}$). 
Further increasing the parameter ($\lambda \ge 600$) leads to a slight increase in $\text{err}_{tot}$, as the penalty formulation also slightly penalizes the tangential velocity component (due to imperfect normal vector $\mathbf{n}_\phi$). 

Consequently, we select $\lambda = 200$ for all subsequent benchmark simulations, as it strongly enforces the impermeability condition $\mathbf{v} \cdot \mathbf{n}_\phi = 0$ at minimal numerical damping of the surface flow.

\subsubsection{Influence of interface thickness}
\begin{figure*}
\centering
    \includegraphics[width=\linewidth]{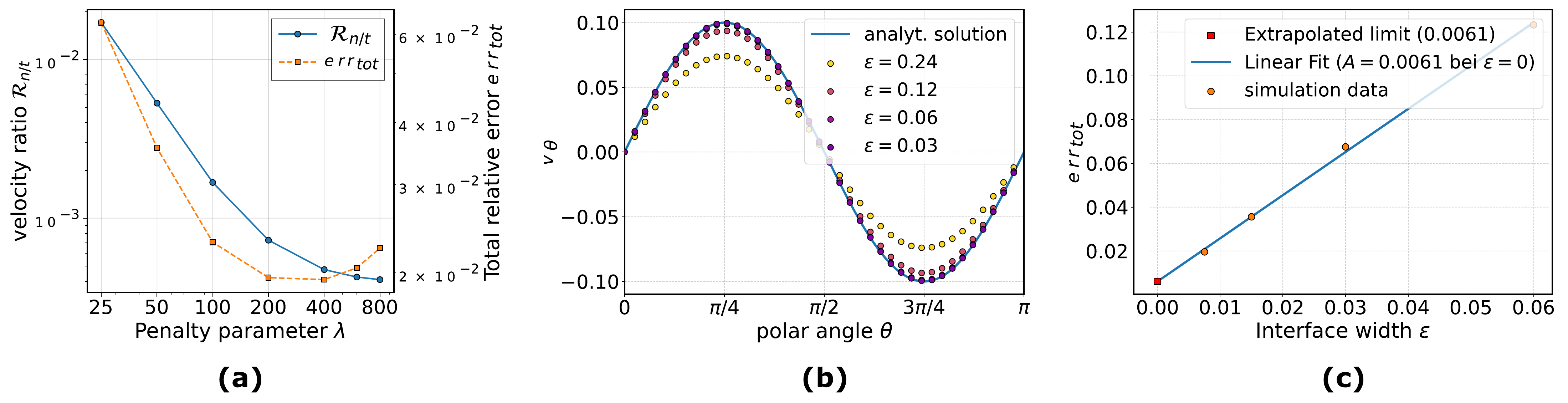}
    \caption{Sensitivity and convergence analysis of the numerical approximation for mode $l=2$. \\
    \textbf{(a):} Influence of the penalty parameter $\lambda$ on the normal-to-tangential velocity ratio $\mathcal{R}_{n/t}$ (blue, left axis) and the total relative error $\text{err}_{tot}$ (orange, right axis). Connecting lines serve merely as a guide to the eye. \textbf{(b):} Tangential velocity profile $v_\theta$ along the polar angle $\theta$, comparing the analytical solution (blue line) against numerical approximations for progressively refined interface widths. 
    \textbf{(c):} Linear convergence of the total relative error $\text{err}_{tot}$ with respect to the interface width $\epsilon$. A Richardson extrapolation towards the sharp-interface limit ($\epsilon \to 0$) yields a residual error of $0.0061$ ($0.61\%$). 
    \textbf{Simulation parameters:} $\eta_b = \eta_s=1$, $\eta_0 = 0.05$, $\eta_1 = 0.95$, $\eta_b=1$, $\epsilon = 0.0075$ (a), $\lambda = 200$ (b,c), $\sigma = 0$, $\Delta \vx = 0.2$}
    \label{fig:validation1}
\end{figure*}

To systematically evaluate the accuracy of the diffuse interface approximation, we compare the obtained tangential velocity of our diffuse interface model with the analytical solution. Fig.~\ref{fig:validation1}b shows numerical tangential velocity profiles $v_{\theta, \phi}$ extracted along the $\phi=0.5$ isoline for various values of the interface width $\epsilon$, compared against the analytical sharp interface solution $v_\theta = \frac{1}{5}\cos(\theta)\sin(\theta)$. To ensure that the diffuse interface remains sufficiently resolved, the mesh size was refined proportionally to the interface thickness, i.e., $\Delta \vx = \mathcal{O}(\epsilon)$, throughout the parameter study.

As the diffuse interface narrows, the numerical approximation converges monotonically towards the analytical solution, becoming visually indistinguishable from the theoretical curve for $\epsilon \le 0.03$ with an error of $\text{err}_{tot}\approx 0.068$. The error of almost $7\%$---despite the visual conformity---stems from stronger deviations in the interface region apart from the $\phi=0.5$ level set. 

To further assess the asymptotic behavior in the sharp interface limit ($\epsilon \to 0$), Fig.~\ref{fig:validation1}c illustrates the total relative $L_2$ error ($\text{err}_{tot}$) as a function of $\epsilon$.
The error exhibits a clear linear dependence on the interface width, consistent with the expected $\mathcal{O}(\epsilon)$ convergence of standard diffuse-interface approximations. Applying a linear Richardson extrapolation to the simulation data yields a residual limiting error of $\text{err}_{tot} \approx 0.0061$ ($0.61\%$), which likely stems from the spatial discretization. 
In conclusion, these results confirm the expected convergence of the diffuse-interface solutions towards the sharp-interface reference solution. 

\subsubsection{Dependence on the Governing Parameters}
In the following we aim to extend the previous comparison to a wider range of parameters, to challenge contributions of the various viscous terms in the model. 
To systematically evaluate these dependencies, we compare the maximum tangential velocity of our numerical simulations evaluated at the $\phi = 0.5$ isoline against the analytical sharp interface predictions. 
This measure allows us to highlight how sensitive the velocity depends on parameters and how well this sensitivity is captured by our simulations.

\begin{figure*}
    \centering  
    \includegraphics[width=\linewidth]{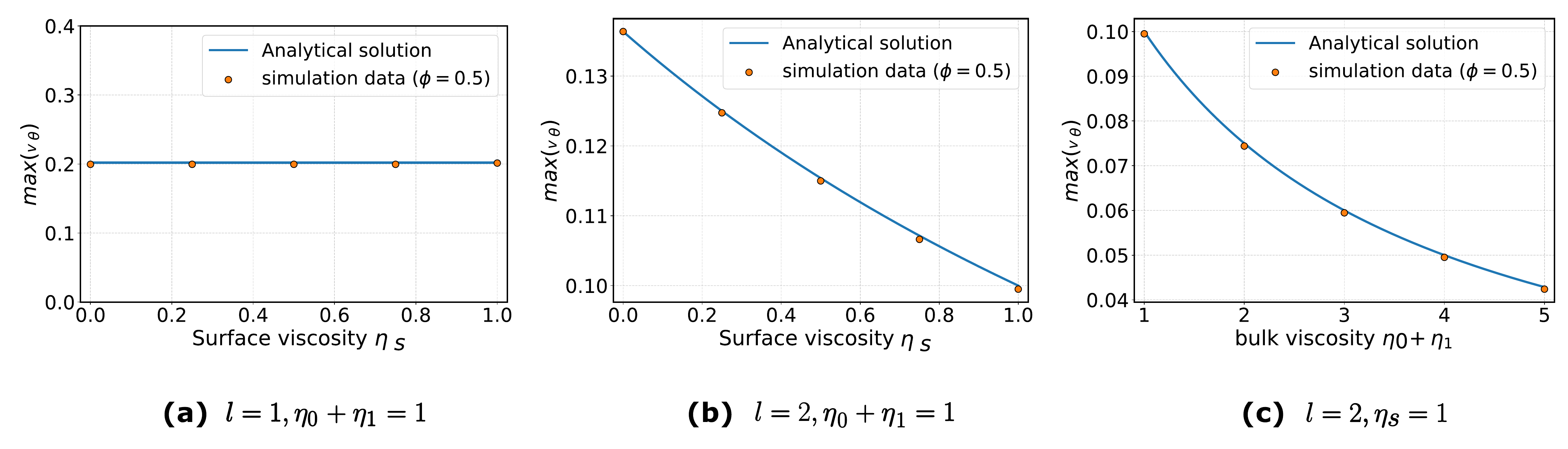}
    \caption{Influence of the governing parameters on tangential velocity shows excellent agreement between numerical results (marker points) and analytical solution (lines, based on Eq.~\eqref{eq: analytical tangential velocity}). 
    \textbf{Further parameters:} 
    $\eta_0/\eta_1=0.05/0.95$, $\eta_b=1$, $\epsilon = 0.0075$, $\lambda = 200$, $\Delta \vx = 0.2$ 
    }
    \label{fig:validation2}
\end{figure*}

First, we investigate the mode $l=1$. According to the analytical model in Eq.~\eqref{eq: analytical tangential velocity}, the denominator for $l=1$ evaluates to $2\eta_0 + 3\eta_1 + 2\eta_b$. 
Consequently, the theoretical flow field is expected to be completely independent of $\eta_s$. This provides a particularly stringent test of the implementation, since an artificial dependence on $\eta_s$ would indicate an inconsistency in the discretized surface viscous operator. Our numerical simulations reproduce this non-trivial property very well: across varying surface viscosities, the maximum tangential velocity remained constant at approximately $0.2$, which equals the sharp interface value, confirming that the numerical framework correctly captures this decoupling, see Fig.~\ref{fig:validation2}a. 

Next, we systematically analyze the flow field for mode $l=2$, where the tangential velocity depends on both $\eta_s$ and $\eta_b$.  
The analytical solution predicts the maximum tangential velocity to scale proportional to $(5(\eta_0+\eta_1) + 6\eta_b + 4\eta_s)^{-1}$. 
The numerical data is in excellent agreement with this theoretical scaling when both $\eta_s$ or $(\eta_0+\eta_1)$ are varied (Figs.~\ref{fig:validation2}b and \ref{fig:validation2}c), which implicitly also validates the $\eta_b$ contribution. 
The diffuse interface model accurately captures the dampening effect of both the shear and dilational surface viscosities on the tangential flow. 
The excellent agreement across all parameter sweeps provides strong validation for the model and confirms its capability to accurately resolve complex interfacial flow dynamics governed by varying viscous and active forces.

\subsection{Topology change} \label{sec:topological change}
Topological changes arise in different real world processes, e.g. during cell division or when two vesicles fuse together. In contrast to grid-based surface descriptions, the phase-field method allows to handle topological changes without any effort. 
In the following, we demonstrate that simulations based on our novel energy-stable discretization remain thermodynamically consistent throughout topological changes. To the best of our knowledge, the preservation of the discrete energy-dissipation law across such topological transitions has not been reported previously for a viscous surface. 

Here, we simulate the fusion of two axisymmetric vesicles with a viscous surface. 
Consistent with the rheology of lipid bilayer membranes, we choose a dominant dilational viscosity ($\eta_b \gg \eta_s > 0$). 
Bending contributions could be easily included in the model~\cite{du2005phase,Aland_2014_inextensibility} but are omitted here. 
The vesicles are initially prescribed by an elliptic shape (axes lengths $1.0$ and $0.5$) and are placed with a distance of $0.1$ to each other in the center of the domain $\Omega = [-4, 4] \times [0, 4]$. A positive surface tension $\sigma = 3$ leads to their evolution towards a circular shape. 
As soon as the interfaces touch each other the fusion process begins and, due to the positive surface tension, the dumbbell-shaped fusion deforms slowly towards a single spherical vesicle.

\begin{figure*}
    \centering
    \includegraphics[width=1\linewidth]{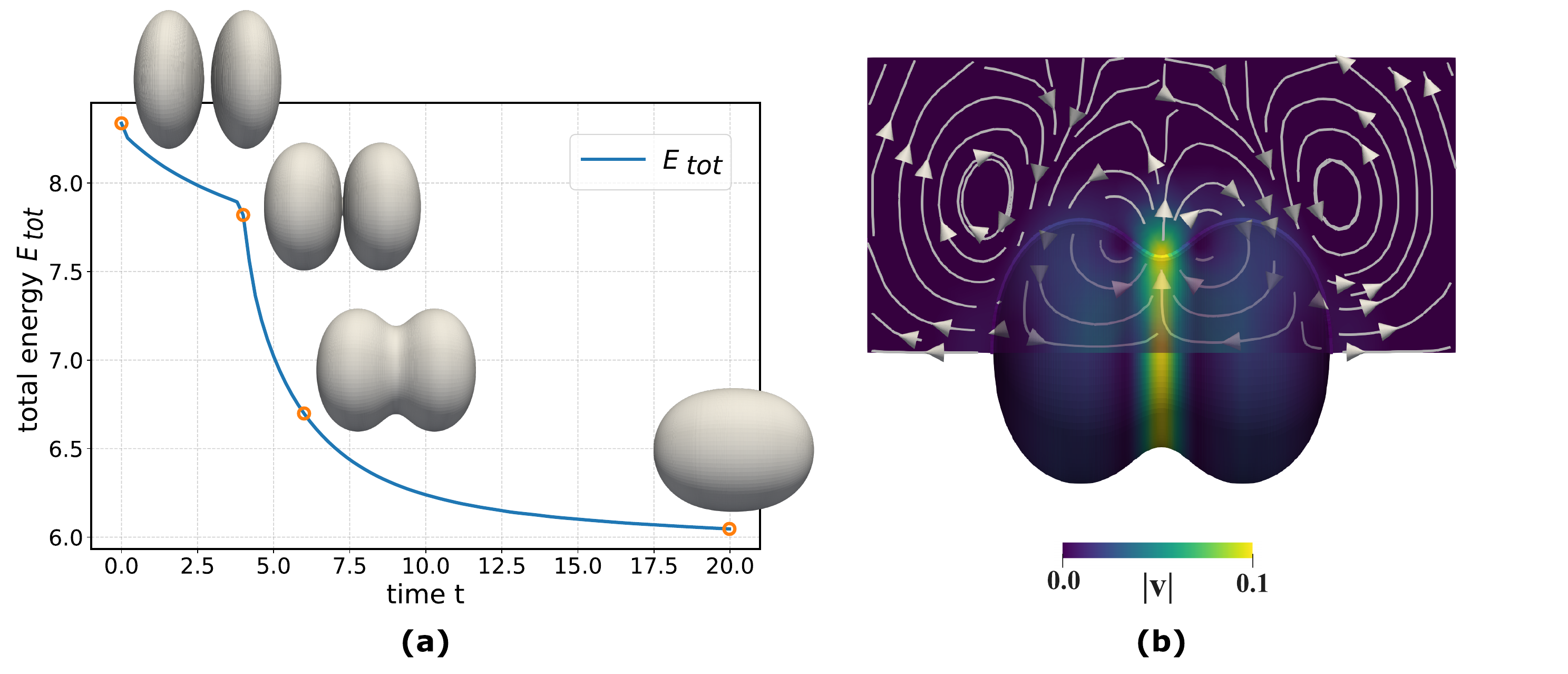}
    \caption{Illustration of discrete energy dissipation for the fusion of two oblate vesicles. \\
    \textbf{(a):} Evolution of the total energy over time. 
    \textbf{(b):} Snapshot of the surface tension driven flow at $t = 6.0$. The surface of the fusing vesicles is colored by the magnitude of the fluid velocity $\vert \vv \vert$.
    \textbf{Simulation parameters:} $\eta_s=0.01, \eta_b = 1$, $\eta_0 = \eta_1 = 0.1$, $\rho_0 = \rho_1 = 1.0$, $\sigma = 3 $, $\epsilon = 0.03$, $M = 10^{-5}$, $\Delta t = 0.02$, $\Delta \vx = 0.4$, $\text{tol} = 10^{-7}$}
    \label{fig:energy dissipation fusion example}
\end{figure*}

Fig.~\ref{fig:energy dissipation fusion example}a shows snapshots of the fusion process for the times $t = 0, 4, 6, 20$, highlighting the characteristic stages of the topological transition from two disconnected vesicles to a single connected domain. As the vesicles initially approach each other the total energy $E_{tot}$ decreases moderately. Upon contact ($t=4$), a pronounced decrease in energy is observed as the two vesicles merge and form a transient dumbbell-shaped configuration ($t = 6$). Fig.~\ref{fig:energy dissipation fusion example}b illustrates the coalescence step further by showing the corresponding velocity magnitude and streamlines. The strongest flow occurs near the developing neck while additionally a recirculating flow is developed around the vesicles. 
Subsequently, the energy keeps decreasing as the fused vesicle relaxes towards the spherical minimal energy state.  

Notably, the monotonic decrease of the total energy is preserved throughout the topological transition, demonstrating that the discrete energy-dissipation property is maintained even during vesicle coalescence.

\section{Conclusion}\label{sec:conclusion}
In this paper, we presented a thermodynamically consistent phase-field model for viscous hyper-surfaces in incompressible Navier--Stokes fluids. The implicit representation of the surface via a phase-field allows us to describe large deformations and topological changes without explicit surface tracking. Viscous surface stresses are incorporated through the classical Boussinesq--Scriven constitutive law, accounting for both surface shear and dilatational viscosity. This work thus extends existing diffuse-interface models for elastic and viscoelastic surfaces~\cite{KLOPPE2024} to interfaces with an intrinsic viscous response.

We validated the modeling approach by matched asymptotic analysis, where we successfully recovered the corresponding sharp-interface equations in the limit of a vanishing interface thickness (Sec.~\ref{sec:matched asymp analysis}).
A central aspect of our continuous model is its thermodynamic consistency (Sec.~\ref{sec: diffuse interface model}), to maintain which we derived a stable, unconditionally energy-dissipative discretization (Sec.~\ref{sec:discretization}). 

To validate the proposed model, we conducted several numerical experiments for both stationary and dynamically evolving viscous interfaces (Secs.~\ref{sec: viscous band}--\ref{sec:topological change}).
First, using a qualitative shear-flow experiment, we showed that increasing surface viscosity suppresses interfacial stretching, rendering the interface  effectively inextensible for high enough surface viscosity (Sec.~\ref{sec: viscous band}).

Second, we considered flow on a rigid spherical interface as a rigorous quantitative validation problem (Sec.~\ref{sec:validation rigid sphere}). The computed velocity profiles and their dependence on the surface viscosity agree very well with the corresponding analytical sharp-interface solutions. 

Finally, we demonstrated the model's capability to handle topological changes by simulating the fusion of two viscous axisymmetric vesicles (Sec.~\ref{sec:topological change}). The simulation captures the successive stages of the topological transition, including the formation and relaxation of the transient neck. Notably, the total energy decreases monotonically throughout the entire transition, illustrating that the discrete energy-dissipation property is preserved also for dynamically evolving interfaces.

Taken together, these results demonstrate that the proposed phase-field framework can account for finite surface shear and dilatational viscosities while retaining a thermodynamically consistent and energy-dissipative structure. At the same time, the diffuse-interface representation naturally accommodates large deformations and topological changes without explicit surface tracking. The model therefore provides a basis for further investigations of viscous surface dynamics in complex geometries and multiphase flows.
Its flexible structure readily accommodates additional physical effects, such as bending stiffness and active surface forces (see \cite{Gertack2026}), enabling simulations of biological membranes and cell surface dynamics in the future. 

\backmatter






\section*{Statements and Declarations}

\subsection*{Funding} 
SA acknowledges support from the German Research Foundation (grant AL1705/6). 
The authors gratefully acknowledge computing time on the high-performance compute clusters at TU Freiberg (funded by DFG grant 397252409) and the NHR Center of TU Dresden, jointly supported by the Federal Ministry of Education and Research and the state governments participating in the NHR (www.nhr-verein.de/unsere-partner).  

\subsection*{Competing interests}
The authors have no competing interests to declare that are relevant to the content of this article.




\begin{appendices}
\begin{strip}
\section{Axisymmetric formulation}\label{sec:appendix Axisym}

Throughout this work, multiple simulations used an axisymmetric setup assuming that both the phase-field and the flow are rotationally symmetric about a fixed axis. In this section we present the axisymmetric strong and weak forms of the numerical scheme 
Eqs.~\eqref{eq:time discr. NSCH 1}--\eqref{eq:time discr. NSCH 4}. 
In particular, we show that testing the strong form with the natural measure $r\,\mathrm dr\,\mathrm dx$ eliminates all but a single, explicit zero-order contribution per tensor from the additional terms that appear in the cylindrical formulation. The axisymmetric strong form terms for the  diffuse viscous surface force have been previously introduced in~\cite[Appendix A.5]{Gertack2026}.

\subsection{Axisymmetric setup}
\label{sec:axisymmetric setup}

Let the domain $\Omega\subset\mathbb R^3$ be invariant under rotation about a fixed axis, which we take w.l.o.g. to be the first Cartesian
coordinate axis, denoted by $x$. Moreover, the distance to this axis is denoted by $r\ge0$, and $\varphi\in[0,2\pi)$ is the azimuthal angle. We write $\ve_x,\ve_r,\ve_\varphi$ for the associated orthonormal basis, with the derivatives
\begin{equation}
    \partial_\varphi \ve_r = \ve_\varphi, \qquad
    \partial_\varphi \ve_\varphi = -\ve_r ,
    \label{eq:basis-derivatives}
\end{equation}
while all other derivatives of $\ve_x,\ve_r,\ve_\varphi$ with respect to
$x,r,\varphi$ equal zero. All scalar fields ($\phi,\mu,p,\ldots$) are
assumed independent of $\varphi$, and the velocity field is purely meridional,
\begin{equation}
    \vv = v_x\,\ve_x + v_r\,\ve_r , \qquad \partial_\varphi\vv = 0,
    \label{eq:no-swirl}
\end{equation}
i.e.,  we exclude swirl. Under this ansatz $\Omega$ is generated by
rotating a meridian domain $\Omega_{\rm 2D}\subset\{(x,r):r\ge0\}\subset\mathbb
R^2$ about the $x$-axis. Also every volume integral reduces in this setting to
\begin{equation}
    \int_\Omega f \,\mathrm d\vx
    = 2\pi\int_{\Omega_{\rm 2D}} r\,f \,\mathrm dr\,\mathrm dx .
    \label{eq:measure-reduction}
\end{equation}
We drop the constant factor $2\pi$ throughout this work, so that every weak form
below is understood to be tested with the measure $r\,\mathrm
dr\,\mathrm dx$ on $\Omega_{\rm 2D}$.

Now, in cylindrical coordinates, the gradient $\nabla_{R}$ and Laplacian $\Delta_{R}$ of a $\varphi$-independent scalar field $f$ and the gradient and divergence of the velocity field \eqref{eq:no-swirl} read
\begin{align}
    \nabla_{R} f &= \partial_x f\,\ve_x + \partial_r f\,\ve_r ,
    \label{eq:grad-scalar}\\
    \Delta_{R} f &= \partial_x^2 f + \partial_r^2 f + \frac1r\partial_r f ,
    \label{eq:ax-laplace} \\
    \nabla_{R}\cdot\vv &= \partial_x v_x + \partial_r v_r + \frac{v_r}{r},
    \label{eq:ax-vec-div} \\
    \nabla_{R}\vv &=
    \sum_{k,l\in\{x,r\}} \partial_k v_l\; \ve_k\otimes\ve_l
    + \frac{v_r}{r}\; \ve_\varphi\otimes\ve_\varphi ,
    \label{eq:grad-vector} 
\end{align}
i.e., in particular $\nabla_{\!R}\vv$ is block diagonal.

For a general second-order tensor field $\vT$ that is block diagonal in
the same sense, i.e.,
\begin{equation}\label{eq: general tensor  T}
    \vT = \begin{pmatrix}
        T_{xx} & T_{xr} & 0 \\
        T_{rx} & T_{rr} & 0 \\
        0 & 0 & T_{\varphi \varphi}
    \end{pmatrix},
\end{equation}
 with $T_{x\varphi}=T_{\varphi x}=T_{r\varphi}
=T_{\varphi r}=0$, the axisymmetric divergence reads:
\begin{equation}
    \nabla_{R}\cdot \vT
    = \nabla\cdot \vT_{\rm 2D} + \frac1r\begin{pmatrix}T_{rx}\\ T_{rr}-T_{\varphi\varphi}\end{pmatrix},
    \label{eq:ax-tensor-div}
\end{equation}
where $\vT_{\rm 2D}$ is the upper-left $\{x,r\}$-block of the tensor $\vT$.
Note that $\vP_\phi^{n-1}$, the viscous stress tensor
$2\eta(\phi^{n-1})\mathbf D(\vv^n)$, and the viscous surface stress tensor $\vS_\phi^n$ in Eq.~\eqref{eq:time discr. NSCH 4} in cylindrical coordinates are exactly of this block-diagonal structure.

\subsection{Axisymmetric strong form}
Using the previous identities and that the $\varphi\varphi$-component of $\vP_{\phi}^{n-1}$ equals $1$, the strong form of the Navier--Stokes--Cahn--Hillard scheme \eqref{eq:time discr. NSCH 1}--\eqref{eq:time discr. NSCH 4} reads:
\begin{align}
     d_t\phi^{n} =& \nabla_R \cdot \left(M \nabla_R \mu^n \right) - \nabla_R \phi^{n-1} \cdot \vv^n, \label{eq: ax NSCH1} \\
     \mu^n =& -\epsilon \left( \Delta_R\phi^n \right) + \epsilon^{-1}\chi(\phi^n,\phi^{n-1}),  \label{eq: ax NSCH2}\\
     0 =&  \nabla_{R}\cdot\vv^n,  \label{eq: ax NSCH3}\\
     \rho \left( d_t\vv^n + \vv^{n-1}\cdot\nabla_{R}\vv^n\right) =& - \nabla_R p^n +\nabla \cdot \left(2\eta(\phi^{n-1})\vD_{\rm 2D}(\vv^n)\right) +\frac{\eta(\phi^{n-1})}{r}
       \begin{pmatrix}\partial_rv_x^n+\partial_xv_r^n\\
       2\partial_rv_r^n-\dfrac{2v_r^n}{r}\end{pmatrix} \label{eq: ax NSCH4} \\
       & + \sigma \mu^n \nabla_R \phi^{n-1} + \nabla \cdot\left(\lvert\nabla\phi^{n-1}\rvert\,\vS_{\phi,2D}^n\right)
       +\frac{\lvert\nabla\phi^{n-1}\rvert}{r}
       \begin{pmatrix}S^n_{rx}\\ S^n_{rr}-S^n_{\varphi\varphi}\end{pmatrix}, \nonumber 
\end{align}
with the block-entries of the viscous surface stress tensor $\vS_{\phi}^n$ in cylindrical coordinates being

\begin{align}
    \vS_{\phi,2D}^n =& 
        (\eta_b-\eta_s) \left(\vP_{\phi,2D}^{n-1}:\nabla\vv^n + \dfrac{v_r^n}{r}\right) \vP_{\phi,2D}^{n-1} \nonumber \\
        &+ 2\eta_s\,\vP_{\phi,2D}^{n-1}\vD_{\rm 2D}(\vv^n)\vP_{\phi,2D}^{n-1}, \\
    S^n_{\varphi\varphi} =& 
        (\eta_b-\eta_s) \left(\vP_{\phi,2D}^{n-1}:\nabla\vv^n + \dfrac{v_r^n}{r}\right)+ 2\eta_s\,\frac{v_r^n}{r}.
\end{align}

\subsection{Axisymmetric weak formulation}
When transforming the presented strong form Eqs.~\eqref{eq: ax NSCH1}--\eqref{eq: ax NSCH4} into their weak form, we can make use of the following identity.
\begin{lemma}
\label{lem:cancellation}
Let $\Omega_{\rm 2D}\subset\{(x,r):r\ge0\}$ and let $T:\Omega_{\rm 2D}\to\mathbb
R^{3\times3}$ be block diagonal, with $T_{x\varphi}=T_{\varphi
x}=T_{r\varphi}=T_{\varphi r}=0$, and let $\nabla_{R} \cdot T$ be given
by \eqref{eq:ax-tensor-div}. Then, for every $\vw=(w_x,w_r)$ vanishing
on $\partial\Omega_{\rm 2D}\setminus\{r=0\}$ (or satisfying natural boundary
conditions there), it holds
\begin{equation}
    \int_{\Omega_{\rm 2D}} r \vw \cdot \left(\nabla_{R}\cdot T \right)
    \,\mathrm dr\,\mathrm dx
    = -\int_{\Omega_{\rm 2D}} \sum_{k,l \in \{x,r\}} r\,T_{kl}\,\partial_k w_l \,\mathrm dr\,\mathrm dx
      - \int_{\Omega_{\rm 2D}} T_{\varphi\varphi}\,w_r \,\mathrm dr\,\mathrm dx.
    \label{eq:cancellation-lemma}
\end{equation}
In particular, the terms $T_{rl}/r$, $l \in \{x,r\}$, in \eqref{eq:ax-tensor-div} are cancelled out exactly
when performing integration by parts and therefore do not appear on the right-hand side of \eqref{eq:cancellation-lemma}.
\end{lemma}
To show the validity of Lemma~\ref{eq:cancellation-lemma} we insert \eqref{eq:ax-tensor-div} into the left-hand side of
\eqref{eq:cancellation-lemma} and split the sum over $l\in\{x,r\}$:
\begin{align}
     &\int_{\Omega_{\rm 2D}} r\,w_l\Bigl[\sum_{k}\partial_kT_{kl}
    + \frac1r T_{rl} - \frac1r T_{\varphi\varphi}\delta_{l,r}\Bigr]
    \mathrm dr\,\mathrm dx \nonumber \\
    =& \int_{\Omega_{\rm 2D}} r\,w_l\sum_k\partial_kT_{kl}\,\mathrm dr\,\mathrm dx \nonumber \\
    &+ \int_{\Omega_{\rm 2D}} w_l T_{rl}\,\mathrm dr\,\mathrm dx
    - \int_{\Omega_{\rm 2D}} w_r T_{\varphi\varphi}\,\mathrm dr\,\mathrm dx \label{eq:proof-step1}.
\end{align}
Integrating the first term on the right of \eqref{eq:proof-step1} by
parts in $k$, we note that the boundary terms vanish on $\partial\Omega_{\rm 2D}\setminus\{r=0\}$ by the assumption on $\vw$, and at $r=0$ they vanish because the
integrand carries the explicit factor $r$. We get
\begin{align}
    &\int_{\Omega_{\rm 2D}} \sum_{k} r\,w_l\,\partial_kT_{kl}\,\mathrm dr\,\mathrm dx \nonumber \\
    =& -\int_{\Omega_{\rm 2D}} \sum_{k} \partial_k(r\,w_l)\,T_{kl}\,\mathrm dr\,\mathrm dx \nonumber \\
    =& -\int_{\Omega_{\rm 2D}} \sum_{k} r\,\partial_kw_l\,T_{kl}\,\mathrm dr\,\mathrm dx \nonumber\\
      &- \int_{\Omega_{\rm 2D}} \sum_{k} w_l\,T_{kl}\,(\partial_k r)\,\mathrm dr\,\mathrm dx .
    \label{eq:proof-step2}
\end{align}
Since $\partial_k r=\delta_{k,r}$, the last integral in
\eqref{eq:proof-step2} equals $\int_{\Omega_{\rm 2D}} w_lT_{rl}\,\mathrm
dr\,\mathrm dx$, i.e., exactly the second term on the
right-hand side of \eqref{eq:proof-step1}, with the opposite sign.
Substituting \eqref{eq:proof-step2} back into
\eqref{eq:proof-step1}, the two $T_{rl}$-contributions cancel, and
what remains is \eqref{eq:cancellation-lemma}.

Applying Lemma~\ref{lem:cancellation} now leads to the fully time-discrete axisymmetric weak formulation:
Given $\phi^{n-1}$, $\vv^{n-1}$, find $(\phi^n,\mu^n,\vv^n,p^n)$ such that for all admissible test
functions $(\psi,\xi,q,\vu)$, it holds
\begin{align}
    \int_{\Omega_{\rm 2D}} r\,\psi\,d_t\phi^n \,\mathrm dr\,\mathrm dx
    &= -\int_{\Omega_{\rm 2D}} r\,M\,\nabla_{R}\mu^n\cdot\nabla_{R}\psi
    \,\mathrm dr\,\mathrm dx
    - \int_{\Omega_{\rm 2D}} r\,\psi\,\nabla_{R}\phi^{n-1}\cdot\vv^n
    \,\mathrm dr\,\mathrm dx, \\
    \int_{\Omega_{\rm 2D}} r\,\xi\,\mu^n \,\mathrm dr\,\mathrm dx
    &= \int_{\Omega_{\rm 2D}} r\,\epsilon\,\nabla_{R}\phi^n\cdot\nabla_{R}\xi
    \,\mathrm dr\,\mathrm dx
    + \int_{\Omega_{\rm 2D}} r\,\epsilon^{-1}\,\xi\,
      \chi(\phi^n,\phi^{n-1})\,\mathrm dr\,\mathrm dx, \\
    0 &= \int_{\Omega_{\rm 2D}} r\,q\,\nabla_R \cdot \vv^n \,\mathrm dr\,\mathrm dx, \\
    \int_{\Omega_{\rm 2D}} r \, \rho \, d_t\vv^n \cdot \vu \,\mathrm dr\,\mathrm dx
    &= - \int_{\Omega_{\rm 2D}} r\rho \left(\vv^{n-1}\cdot\nabla_{R}\vv^n
    \right) \cdot \vu \,\mathrm dr\,\mathrm dx \\
    &\quad + \int_{\Omega_{\rm 2D}} r\, p^n\,\nabla_R \cdot \vu\,\mathrm dr\,\mathrm dx \notag \\
    &\quad -2\int_{\Omega_{\rm 2D}} r \eta(\phi^{n-1})\,\mathbf D_{\rm 2D}(\vv^n) : 
      \nabla\vu \,\mathrm dr\,\mathrm dx \notag \\
    &\quad- \int_{\Omega_{\rm 2D}} 2\eta(\phi^{n-1})\frac{v_r^n\,u_r}{r}
      \,\mathrm dr\,\mathrm dx \notag\\
    &\quad + \int_{\Omega_{\rm 2D}} r\,\sigma\,\mu^n\,\nabla_{\!R}\phi^{n-1}
      \cdot\vu \,\mathrm dr\,\mathrm dx \notag\\
    &\quad -\int_{\Omega_{\rm 2D}} r\,(\eta_b-\eta_s)\,\lvert\nabla\phi^{n-1}\rvert \, (\vP_{\phi,2D}^{n-1}:\nabla \vv^n)\,(\vP_{\phi,2D}^{n-1}:\nabla \vu) \,\mathrm dr\,\mathrm dx \notag \\ 
    &\quad -\int_{\Omega_{\rm 2D}} (\eta_b-\eta_s)\,\lvert\nabla\phi^{n-1}\rvert \left( u_r (\vP_{\phi,2D}^{n-1}:\nabla \vv^n) + v_r^n (\vP_{\phi,2D}^{n-1}:\nabla \vu) \right) \,\mathrm dr\,\mathrm dx \notag \\
    &\quad -\int_{\Omega_{\rm 2D}} (\eta_b-\eta_s)\,\lvert\nabla\phi^{n-1}\rvert \frac{v_r^n u_r}{r} \,\mathrm dr\,\mathrm dx \notag \\
    &\quad -\int_{\Omega_{\rm 2D}} \sum_{k,l \in \{x,r\}} r \eta_s\,\lvert\nabla\phi^{n-1}\rvert 
      \left[\vP_{\phi,2D}^{n-1}(\nabla\vv^n+(\nabla\vv^n)^T)
      \vP_{\phi,2D}^{n-1}\right]_{kl}\,\partial_k u_l
      \,\mathrm dr\,\mathrm dx \notag \\
    &\quad - \int_{\Omega_{\rm 2D}} 2\eta_s\,\lvert\nabla\phi^{n-1}\rvert
      \frac{v_r^n\,u_r}{r} \,\mathrm dr\,\mathrm dx \notag.
\end{align}    
In line with Lemma~\ref{lem:cancellation}, every $1/r$-weighted
meridional correction vector of the strong momentum balance
\eqref{eq: ax NSCH4} is reproduced exactly by the integration by parts, s.t. the weak formulation above retains, for each of the two
tensor-divergence terms (viscous bulk stress and viscous surface stress), only a single explicit zero-order contribution of the form $v_r^nu_r/r$.
\end{strip}

\end{appendices}


\bibliography{literature}

@article{Sagis2011,
   author = {Leonard M C Sagis},
   doi = {10.1103/RevModPhys.83.1367},
   issue = {4},
   journal = {Reviews of Modern Physics},
   pages = {1367},
   title = {Dynamic properties of interfaces in soft matter: Experiments and theory},
   volume = {83},
   year = {2011}
}

@article{demont2026consistency,
  title={On the consistency of dynamic wetting boundary conditions for the Navier--Stokes--Cahn--Hilliard equations},
  author={Demont, T.H.B. and Stoter, Stein K.F and Diddens, Christian and van Brummelen, E.H},
  journal={Computer Methods in Applied Mechanics and Engineering},
  volume={449},
  pages={118519},
  year={2026},
  doi = {https://doi.org/10.1016/j.cma.2025.118519},
  publisher={Elsevier}
}

@article{qian2006variational,
  title={A variational approach to moving contact line hydrodynamics},
  author={Qian, Tiezheng and Wang, Xiao-Ping and Sheng, Ping},
  journal={Journal of Fluid Mechanics},
  volume={564},
  pages={333--360},
  year={2006},
  doi={10.1017/S0022112006001935},
  publisher={Cambridge University Press}
}

@article{kovacs2024numerical,
  title={Numerical surgery for mean curvature flow of surfaces},
  author={Kov{\'a}cs, Bal{\'a}zs},
  journal={SIAM Journal on Scientific Computing},
  volume={46},
  number={2},
  pages={A645--A669},
  year={2024},
  doi = {10.1137/22M1531919},
  publisher={SIAM}
}

@article{sahu2020arbitrary,
  title={Arbitrary Lagrangian--Eulerian finite element method for curved and deforming surfaces: I. General theory and application to fluid interfaces},
  author={Sahu, Amaresh and Omar, Yannick A.D and Sauer, Roger A and Mandadapu, Kranthi K},
  journal={Journal of Computational Physics},
  volume={407},
  pages={109253},
  year={2020},
  publisher={Elsevier},
  doi = {https://doi.org/10.1016/j.jcp.2020.109253}
}

@article{du2005phase,
  title={A phase field formulation of the Willmore problem},
  author={Du, Qiang and Liu, Chun and Ryham, Rolf and Wang, Xiaoqiang},
  journal={Nonlinearity},
  volume={18},
  number={3},
  pages={1249--1267},
  doi={10.1088/0951-7715/18/3/016},
  year={2005}
}

@article{Demont2023,
    title={Numerical investigation of the sharp-interface limit of the Navier–Stokes–Cahn–Hilliard equations}, 
    volume={970}, 
    doi={10.1017/jfm.2023.611}, 
    journal={Journal of Fluid Mechanics}, 
    author={Demont, T.H.B. and Stoter, S.K.F. and van Brummelen, E.H.}, 
    year={2023}, 
    pages={A24}
}

@article{magaletti2013sharp,
  title={The sharp-interface limit of the Cahn--Hilliard/Navier--Stokes model for binary fluids},
  author={Magaletti, Francesco and Picano, Francesco and Chinappi, Mauro and Marino, Luca and Casciola, Carlo Massimo},
  journal={Journal of Fluid Mechanics},
  volume={714},
  pages={95--126},
  year={2013},
  doi={10.1017/jfm.2012.461},
  publisher={Cambridge University Press}
}

@article{abels2014sharp,
  title={On sharp interface limits for diffuse interface models for two-phase flows},
  author={Abels, Helmut and Lengeler, Daniel},
  journal={Interfaces and Free Boundaries},
  volume={16},
  number={3},
  pages={395--418},
  doi={10.4171/IFB/324},
  year={2014}
}

@article{KLOPPE2024,
title = {A phase-field model of elastic and viscoelastic surfaces in fluids},
journal = {Computer Methods in Applied Mechanics and Engineering},
volume = {428},
pages = {117090},
year = {2024},
issn = {0045-7825},
doi = {https://doi.org/10.1016/j.cma.2024.117090},
url = {https://www.sciencedirect.com/science/article/pii/S0045782524003463},
author = {Maximilian Kloppe and Sebastian Aland}
}

@article{grun2014two,
  title={Two-phase flow with mass density contrast: stable schemes for a thermodynamic consistent and frame-indifferent diffuse-interface model},
  author={Gr{\"u}n, G{\"u}nther and Klingbeil, Fabian},
  journal={Journal of Computational Physics},
  volume={257},
  pages={708--725},
  year={2014},
  doi = {https://doi.org/10.1016/j.jcp.2013.10.028},
  publisher={Elsevier}
}

@article{ALAND2014_TimeIntegration,
title = {Time integration for diffuse interface models for two-phase flow},
journal = {Journal of Computational Physics},
volume = {262},
pages = {58-71},
year = {2014},
issn = {0021-9991},
doi = {https://doi.org/10.1016/j.jcp.2013.12.055},
url = {https://www.sciencedirect.com/science/article/pii/S0021999114000102},
author = {Sebastian Aland}
}

@misc{PraetoriusAMDiS,
  author = {Praetorius, S.},
  title = {The Adaptive Multi-Dimensional Simulation Toolbox ({AMDiS}), a discretization module on top of the {Dune}
framework},
  howpublished = {\url{https://gitlab.com/amdis/amdis}}
}

@article{AMDIS_2007,
author = {Vey, Simon and Voigt, Axel},
title = {{AMDiS}: adaptive multidimensional simulations},
journal = {Computing and Visualization in Science},
volume = {10},
pages = {57-67},
year = {2007},
doi = {10.1007/s00791-006-0048-3},
URL = { 
        https://doi.org/10.1007/s00791-006-0048-3
}
}

@article{Witkowski_adaptive_parallel_FEM,
title = {Software concepts and numerical algorithms for a scalable adaptive parallel finite element method},
journal = {Advances in Computational Mathematics},
volume = {41},
pages = {1145--1177},
year = {2015},
issn = {1572-9044},
doi = {https://doi.org/10.1007/s10444-015-9405-4},
author = {Witkowski, T. and Ling, S. and Praetorius, S. and Voigt, A.}
}

@article{Davis_UMFPACK,
author = {Davis, Timothy A.},
title = {Algorithm 832: UMFPACK V4.3---an Unsymmetric-Pattern Multifrontal Method},
year = {2004},
issue_date = {June 2004},
publisher = {Association for Computing Machinery},
address = {New York, NY, USA},
volume = {30},
number = {2},
issn = {0098-3500},
url = {https://doi.org/10.1145/992200.992206},
doi = {10.1145/992200.992206},
journal = {ACM Trans. Math. Softw.},
pages = {196–199}
}

@article{AGG12,
  title={Thermodynamically consistent, frame indifferent diffuse interface models for incompressible two-phase flows with different densities},
  author={Abels, Helmut and Garcke, Harald and Gr{\"u}n, G{\"u}nther},
  journal={Mathematical Models and Methods in Applied Sciences},
  volume={22},
  number={03},
  pages={1150013},
  year={2012},
  doi={https://doi.org/10.1142/S0218202511500138},
  publisher={World Scientific}
}

@phdthesis{mietke2018dynamics,
  title={Dynamics of active surfaces},
  author={Mietke, Alexander},
  year={2018},
  school={Technische Universit{\"a}t Dresden}
}

@article{mietke_min_model19,
  title = {Minimal Model of Cellular Symmetry Breaking},
  author = {Mietke, Alexander and Jemseena, V. and Kumar, K. Vijay and Sbalzarini, Ivo F. and J\"ulicher, Frank},
  journal = {Phys. Rev. Lett.},
  volume = {123},
  issue = {18},
  pages = {188101},
  numpages = {6},
  year = {2019},
  month = {Oct},
  publisher = {American Physical Society},
  doi = {10.1103/PhysRevLett.123.188101},
  url = {https://link.aps.org/doi/10.1103/PhysRevLett.123.188101}
}

@article{MOKBEL2018823,
	title = {A phase-field model for fluid–structure interaction},
	journal = {Journal of Computational Physics},
	volume = {372},
	pages = {823-840},
	year = {2018},
	issn = {0021-9991},
	doi = {https://doi.org/10.1016/j.jcp.2018.06.063},
	url = {https://www.sciencedirect.com/science/article/pii/S0021999118304418},
	author = {Dominic Mokbel and Helmut Abels and Sebastian Aland}
}

@article{scriven1960dynamics,
  title={Dynamics of a fluid interface equation of motion for Newtonian surface fluids},
  author={Scriven, Laurence E},
  journal={Chemical Engineering Science},
  volume={12},
  number={2},
  pages={98--108},
  year={1960},
  publisher={Elsevier},
  doi = {https://doi.org/10.1016/0009-2509(60)87003-0}
}

@article{bothe2010,
  title={On the two-phase Navier--Stokes equations with Boussinesq--Scriven surface fluid},
  author={Bothe, Dieter and Pr{\"u}ss, Jan},
  journal={Journal of Mathematical Fluid Mechanics},
  volume={12},
  number={1},
  pages={133--150},
  year={2010},
  doi={https://doi.org/10.1007/s00021-008-0278-x},
  publisher={Springer}
}

@article{Barrett2015,
  author  = {Barrett, John W. and Garcke, Harald and N{\"u}rnberg, Robert},
  title   = {Stable Numerical Approximation of Two-Phase Flow with a {Boussinesq--Scriven} Surface Fluid},
  journal = {Communications in Mathematical Sciences},
  volume  = {13},
  number  = {7},
  pages   = {1829--1874},
  year    = {2015},
  doi     = {10.4310/CMS.2015.v13.n7.a9}
}

@article{den2007intermonolayer,
  title={Intermonolayer friction and surface shear viscosity of lipid bilayer membranes},
  author={{den Otter}, Wouter K and Shkulipa, S.A},
  journal={Biophysical journal},
  volume={93},
  number={2},
  pages={423--433},
  year={2007},
  doi = {https://doi.org/10.1529/biophysj.107.105395},
  publisher={Elsevier}
}

@article{LOPEZ1998231,
title = {Direct Determination of the Dependence of the Surface Shear and Dilatational Viscosities on the Thermodynamic State of the Interface: Theoretical Foundations},
journal = {Journal of Colloid and Interface Science},
volume = {206},
number = {1},
pages = {231-239},
year = {1998},
issn = {0021-9797},
doi = {https://doi.org/10.1006/jcis.1998.5720},
url = {https://www.sciencedirect.com/science/article/pii/S0021979798957204},
author = {J.M. Lopez and A. Hirsa}
}

@article{MENDOZA2014303,
title = {Particle laden fluid interfaces: Dynamics and interfacial rheology},
journal = {Advances in Colloid and Interface Science},
volume = {206},
pages = {303-319},
year = {2014},
issn = {0001-8686},
doi = {https://doi.org/10.1016/j.cis.2013.10.010},
url = {https://www.sciencedirect.com/science/article/pii/S0001868613001255},
author = {Alma J. Mendoza and Eduardo Guzmán and Fernando Martínez-Pedrero and Hernán Ritacco and Ramón G. Rubio and Francisco Ortega and Victor M. Starov and Reinhard Miller}
}

@article{LiZhang2021,
  author  = {Li, Ping and Zhang, Junfeng},
  title   = {Similar but Distinct Roles of Membrane and Interior Fluid Viscosities in Capsule Dynamics in Shear Flows},
  journal = {Cardiovascular Engineering and Technology},
  volume  = {12},
  number  = {2},
  pages   = {232--249},
  year    = {2021},
  doi     = {10.1007/s13239-020-00517-4}
}

@article{WITTWER2023100126,
title = {A computational model of self-organized shape dynamics of active surfaces in fluids},
journal = {Journal of Computational Physics: X},
volume = {17},
pages = {100126},
year = {2023},
issn = {2590-0552},
doi = {https://doi.org/10.1016/j.jcpx.2023.100126},
url = {https://www.sciencedirect.com/science/article/pii/S2590055223000045},
author = {Lucas D. Wittwer and Sebastian Aland}
}

@article{DEKINKELDER2021110413,
	title = {A numerical method for the simulation of viscoelastic fluid surfaces},
	journal = {Journal of Computational Physics},
	volume = {440},
	pages = {110413},
	year = {2021},
	issn = {0021-9991},
	doi = {https://doi.org/10.1016/j.jcp.2021.110413},
	url = {https://www.sciencedirect.com/science/article/pii/S0021999121003089},
	author = {Eloy {de Kinkelder} and Leonard Sagis and Sebastian Aland}
}

@article{shen2018scalar,
  title={The scalar auxiliary variable ({SAV}) approach for gradient flows},
  author={Shen, Jie and Xu, Jie and Yang, Jiang},
  journal={Journal of Computational Physics},
  volume={353},
  pages={407--416},
  year={2018},
  doi = {https://doi.org/10.1016/j.jcp.2017.10.021},
  publisher={Elsevier}
}

@article{aland2016efficient,
  title={An efficient and energy stable scheme for a phase-field model for the moving contact line problem},
  author={Aland, Sebastian and Chen, Feng},
  journal={International journal for numerical methods in fluids},
  volume={81},
  number={11},
  pages={657--671},
  doi={https://doi.org/10.1002/fld.4200},
  year={2016}
}

@article{AlandMokbel2021,
author = {Aland, Sebastian and Mokbel, Dominic},
title = {A unified numerical model for wetting of soft substrates},
journal = {International Journal for Numerical Methods in Engineering},
volume = {122},
number = {4},
pages = {903-918},
doi = {https://doi.org/10.1002/nme.6567},
url = {https://onlinelibrary.wiley.com/doi/abs/10.1002/nme.6567},
year = {2021}
}

@article{Mokbel2024Wetting,
author = {Mokbel, Marcel and Mokbel, Dominic and Liese, Susanne and Weber, Christoph and Aland, Sebastian},
title = {A Simulation Method for the Wetting Dynamics of Liquid Droplets on Deformable Membranes},
journal = {SIAM Journal on Scientific Computing},
volume = {46},
number = {6},
pages = {B806-B829},
year = {2024},
doi = {10.1137/24M1641142},
URL = {    
        https://doi.org/10.1137/24M1641142 
}
}

@article{Aland_2014_inextensibility,
title = {Diffuse interface models of locally inextensible vesicles in a viscous fluid},
journal = {Journal of Computational Physics},
volume = {277},
pages = {32-47},
year = {2014},
issn = {0021-9991},
doi = {https://doi.org/10.1016/j.jcp.2014.08.016},
url = {https://www.sciencedirect.com/science/article/pii/S0021999114005622},
author = {Sebastian Aland and Sabine Egerer and John Lowengrub and Axel Voigt}
}

@article{Gertack2026,
    author = {Gertack, Hanna Luise and Hampshire, Peter A. E. and Wohlgemuth, Claudia and Alert, Ricard and Aland, Sebastian},
    title = {Modes of mechanical guidance of adhesion-independent cell migration},
    journal = {Soft Matter},
    volume = {22},
    number = {4},
    pages = {907-925},
    year = {2026},
    month = {01},
    issn = {1744-683X},
    doi = {10.1039/d5sm00960j},
    url = {https://doi.org/10.1039/d5sm00960j}
}

@article{Park2026,
title = {Diffuse-interface modeling of two-phase flows with a Boussinesq-Scriven interface},
journal = {Computers \& Fluids},
volume = {307},
pages = {106970},
year = {2026},
issn = {0045-7930},
doi = {https://doi.org/10.1016/j.compfluid.2026.106970},
author = {Jang {Min Park}}
}

\end{document}